\ifdefined\pdfoutput
  \pdfoutput=1
\fi
\documentclass[10pt]{article}

\usepackage[a4paper,margin=2.10cm]{geometry}
\usepackage[T1]{fontenc}
\usepackage{lmodern}
\usepackage{float}
\usepackage{needspace}
\usepackage{microtype}
\usepackage{amsmath,amssymb,bm,mathtools}
\usepackage{xcolor}
\usepackage{graphicx}
\usepackage{tikz}
\usetikzlibrary{arrows.meta,positioning,calc,fit,shapes.geometric}
\usepackage{hyperref}
\hypersetup{
  colorlinks=true,
  linkcolor=blue!45!black,
  citecolor=blue!45!black,
  urlcolor=blue!45!black,
  pdftitle={Connecting Heavy-Quarkonium Born--Oppenheimer EFT to the Peskin OPE},
  pdfauthor={Arkadiy I. Syamtomov}
}
\newcommand{\E}{\bm{\mathcal E}}
\newcommand{\dd}{\mathrm d}
\newcommand{\Tr}{\mathrm{Tr}}

\title{\bfseries QCD-Matched Gluonic Response in Heavy-Quarkonium Born–Oppenheimer EFT: Locality, Channel Factorization, and the Peskin Limit}
\author{Arkadiy I. Syamtomov\\[-0.25em]
\small Bogolyubov Institute for Theoretical Physics, National Academy of Sciences of Ukraine}
\date{}

\begin{document}
\maketitle
\vspace{-1.4em}
\begin{abstract}
We formulate a source-dependent QCD-to-BOEFT matching for the
gluonic response of a stable compact heavy-quarkonium state.
The matched complementary-space resolvent preserves coupled
Born--Oppenheimer dynamics, while an exact Feshbach decomposition
separates channels retained explicitly from sectors integrated out
at the next matching step.  Spectral separation, kernel analyticity,
and joint propagation--source bounds provide sufficient conditions
for a local channel-factorized OPE; otherwise low-energy BO poles
and cuts remain dynamical.

The weak-coupling pNRQCD response is recovered as a limiting
reference problem.  With the additional leading-$E1$,
large-$N_c$/free-octet assumptions, the source-weighted spectral
measure reproduces the established Bhanot--Peskin electric moments
and dissociation cut.  For a Coulombic spin-singlet $1S$ state we
also obtain the sequential-$M1$ contribution
$c_{B,GG}^{(1)ij}
=5\pi\alpha_s^2(c_FV_{\rm iso}^{(s)})^2\delta^{ij}/16$
and the covariant-kinetic seagull contribution
$c_{B,\mathrm{dia}}^{(1)ij}
=-\pi\alpha_s^2\delta^{ij}/4$.
These are identifiable components, not the complete magnetic
matching coefficient.
\end{abstract}

\section{Introduction}

Peskin's short-distance construction organizes the forward interaction
of a sufficiently small heavy quarkonium with soft gluonic fields into
local operators~\cite{Peskin1979,BhanotPeskin1979}.  The contemporaneous
multipole analysis of nonperturbative gluonic effects was developed by
Voloshin~\cite{Voloshin1979}, and the QCD multipole expansion was
subsequently reviewed in Ref.~\cite{Kuang2006}.  The physical mechanism
underlying the leading chromoelectric response is a transition from a
color-singlet heavy pair to a complementary color-excited sector,
propagation in that sector, and a second $E1$ transition back to the
singlet.  Potential NRQCD makes the corresponding octet Green function
explicit~\cite{pNRQCD2000,gWEFT2016}.  Effective
quarkonium--gluon Lagrangians were developed in
Ref.~\cite{Luke1992}, while the leading-twist relation to hadronic
scattering was studied in Refs.~\cite{Arleo2001,Arleo2005}.  The
pNRQCD relation between singlet-to-octet breakup and
gluo-dissociation, including the octet final-state interaction, was
established in Ref.~\cite{Brambilla2011}.

Born--Oppenheimer EFT (BOEFT) supplies coupled Schr\"odinger
Hamiltonians for gauge-invariant heavy-pair states carrying nontrivial
light-field quantum numbers
\cite{Berwein2015,BrambillaBO2018,BOEFT2024}.  Its coupled-channel
realization also accommodates quarkonium--open-flavor mixing and the
associated bound-state and resonance poles~\cite{OpenFlavorBO2026}.
A direct precursor to the present analysis is the Born--Oppenheimer
study of Ref.~\cite{LakhinaSwanson2004}, which expressed the
chromoelectric polarizability as a sum over intermediate hybrid states,
identified the dependence of their energy denominators on the BO
quantum numbers as an obstruction to Peskin factorization, and
recovered the Peskin result in the asymptotically heavy-quark limit.
Thus the conceptual relation between BO intermediate states and the
Peskin OPE is established.  The problem addressed here is to formulate
that relation for a general coupled BOEFT response with
gauge-covariantly matched sources, explicitly retained channels, and
irreducible two-field matching contributions.

External-field Green functions have been used in pNRQCD matching, for
example for electromagnetic $E1$ transitions~\cite{BrambillaE12012},
while chromoelectric and chromomagnetic insertions have been matched in
weak-coupling hybrid-to-quarkonium
transitions~\cite{BrambillaTransitions2023}.  Normalized
source-dependent residues and a channel-complete pNRQCD response were
developed in Ref.~\cite{ResonancePolarizability2026}.  The construction
below instead assumes a normalized stable zero-field eigenstate and
implements its response on the coupled BOEFT Hilbert space.  For an
isolated resonance, the response must be defined at the complex pole
with the derivative normalization of
Ref.~\cite{ResonancePolarizability2026}, rather than by the
Hilbert-space projector used here.  In the magnetic sector, the $M1$
singlet--octet dissociation mechanism was calculated in
Ref.~\cite{ChenHe2017}, while its spin-resolved pNRQCD source and the
additional symmetry-allowed one-field tensors were formulated in
Ref.~\cite{YangYao2024}.

Here we define the source-dependent gluonic response on the
gauge-covariant static-source space of stable coupled BOEFT states.
The complementary-space projected resolvent preserves the full matched
channel dynamics.  A subsequent Feshbach decomposition separates
channels retained dynamically from eliminated-sector propagation,
while a separately matched kernel collects contributions irreducible with
respect to a single complementary-sector propagator.  We give sufficient operator-norm
bounds that jointly control light-channel propagation and source
creation; their parametric smallness must be supplied by the EFT power
counting appropriate to the matching problem.  Together with
multipole locality, a source gap, and analyticity of the irreducible
kernel, these bounds yield a channel-factorized local OPE.  The strict Coulombic, leading-$E1$, $V_A=1$,
large-$N_c$/free-octet, and forward on-shell gluon projection then
recovers the Bhanot--Peskin moments and dissociation cut as inverse
moments and a boundary-value discontinuity of the same source-weighted
spectral measure.

As a magnetic extension, we take the subthreshold inverse moment of
the established isotropic $M1$ spectral strength and organize the
radiative scalar and tensor source contributions.  We also isolate the
color-singlet covariant-kinetic diamagnetic term, whose normalization
follows from the Poincar\'e-fixed kinetic operator and the neutral-pair
pseudomomentum construction
\cite{BrambillaPoincare2003,AlfordStrickland2013}.  The contribution of the present work lies in the common
source-normalized, channel-complete organization of these electric and
magnetic responses and in the controlled reduction to the Peskin
limit.  The established $E1$ mechanism, BO hybrid sum,
Bhanot--Peskin moments, and $M1$ dissociation strength serve as inputs
and limiting checks of this construction.

\section{From the BOEFT response to the Peskin OPE}
\subsection{Gauge-covariant source response}

The general response developed below is a BOEFT construction.
We begin with the weak-coupling pNRQCD singlet--octet response only
to fix the short-distance source normalization and the resolvent
convention that the BOEFT construction must reproduce in its
perturbative limit. We absorb the gauge coupling into the chromoelectric field, $\mathcal E_i^a\equiv gE_i^a$.  At leading order in the multipole expansion, the relevant pNRQCD terms are
\begin{align}
\mathcal L={}&S^\dagger(i\partial_0-h_s)S
 +O^{a\dagger}(iD_0-h_o)O^a \notag\\
&+V_A(r;\mu_s)\sqrt{\frac{T_F}{N_c}}
 \left(O^{a\dagger}\,\bm r\!\cdot\!\bm{\mathcal E}^{a}S+\text{h.c.}\right)+\cdots .
\label{eq:pnrqcd}
\end{align}
For a singlet eigenstate $h_s|\phi\rangle=E_\phi|\phi\rangle$, elimination of the octet gives the causal dynamical response
with $\Delta_o=h_o-E_\phi$,
\begin{align}
\alpha_{o}^{ij}(\omega)={}&\frac{T_F}{N_c}
\bigg\langle\phi\bigg|r^iV_A^\dagger
\frac{1}{\Delta_o-\omega-i0}V_A r^j \notag\\[-0.2em]
&\hspace{8mm}+r^jV_A^\dagger
\frac{1}{\Delta_o+\omega+i0}V_A r^i\bigg|\phi\bigg\rangle .
\label{eq:octet-response}
\end{align}
For an isotropic $S$ wave, $\alpha_o^{ij}=\delta^{ij}\alpha_o$ and the scalar coefficient contains the usual factor $1/3$~\cite{gWEFT2016}.

Our aim is to preserve the resolvent structure of
Eq.~\eqref{eq:octet-response} while replacing the perturbative octet
sector by the full matched complementary BOEFT sector and defining its
source by gauge-covariant QCD matching.  Let $H_0$ be the full matched zero-field Hamiltonian within the chosen resolved BO, hybrid, and open-flavor EFT, and let $|\phi\rangle$ be a normalized stable eigenstate, $H_0|\phi\rangle=E_\phi|\phi\rangle$.  Define
\begin{equation}
P_\phi=|\phi\rangle\langle\phi|,\qquad
Q_\phi=1-P_\phi .
\label{eq:projectors}
\end{equation}
All zero-field nonadiabatic and threshold mixing is therefore included before the response is formed.

The QCD-to-BOEFT matching is stated in a classical background connection $\bar A_\mu$.  For an admissible background, parallel transport to a common base point $x_\star=(t_\star,\bm R)$ defines
\begin{equation}
\begin{aligned}
\bar{\mathcal E}_i^a(x;x_\star)&=2\,\Tr\!\left[T^a
{\mathcal W}_{\bar A}(x_\star,x)\,g\bar F_{0i}(x)
{\mathcal W}_{\bar A}(x,x_\star)\right],
\qquad \Tr(T^aT^b)=\frac{\delta^{ab}}2 ,\\[-0.2em]
\bar{\mathcal B}_i^a(x;x_\star)&=-\epsilon_{ijk}\,\Tr\!\left[T^a
{\mathcal W}_{\bar A}(x_\star,x)\,g\bar F_{jk}(x)
{\mathcal W}_{\bar A}(x,x_\star)\right].
\end{aligned}
\label{eq:transported-background}
\end{equation}
Coefficient extraction uses smooth finite-parameter families of the
classical connection,
$\bar A_\mu(\boldsymbol\eta)$, with
$\bar A_\mu(\boldsymbol 0)=0$.  We call a family admissible when its
source derivatives are well defined in the zero-background theory and
the background is of compact temporal support, or is equivalently
defined with the usual adiabatic switching.  All Wilson-line contours
and static-source connectors are fixed as part of the matching
convention and are held fixed when the source derivatives are taken.
For the frequency expansion below, transport between time slices is
along the static worldline at fixed $\bm R$; any additional fixed
spatial connector belongs to the definition of the interpolator.

Heavy-pair and BO interpolators carry the corresponding Wilson lines
to $x_\star$.  Before the final singlet contraction, let
${\mathcal O}_A^c$ denote the light-field component of a complementary
BO interpolator with an adjoint endpoint at $x_\star$.  It transforms
as
${\mathcal O}_A^c\to
D_{\rm adj}^{cd}(x_\star){\mathcal O}_A^d$,
while the transported background in
Eq.~\eqref{eq:transported-background} transforms in the same
fiber~\cite{BOEFT2024}.

Motivated by static-source transfer-matrix constructions involving
adjoint sources~\cite{Philipsen2002,PhilipsenWagner2014}, we represent
the open endpoint in an auxiliary gauge-covariant Hilbert space.  At fixed heavy geometry $(\bm R,\bm r)$, let $\mathcal K_{\rm adj}$ be the completion of light-field wave functionals $\Psi^c[A,q]$ satisfying Gauss' law with one adjoint endpoint at $x_\star$.  Residual gauge transformations act as $\Psi^c\to D_{\rm adj}^{cd}(x_\star)\Psi^d$, and
\begin{equation}
(\Psi,\Xi)_{\rm adj}=\sum_{c=1}^{N_c^2-1}
\int\dd\mu_{\rm kin}\;\Psi^{c*}[A,q]\Xi^c[A,q]
\label{eq:adjoint-inner-product}
\end{equation}
is invariant; $\dd\mu_{\rm kin}$ is the gauge-invariant kinematical measure on the Gauss-law domain.  The static transfer Hamiltonian is self-adjoint with respect to this inner product.  Parallel transport between time slices is supplied by
\begin{equation}
\Phi_{\rm adj}^{cd}(t_2,t_1;x_\star)=
\left[{\mathcal P}\exp\!\left(ig\!\int_{t_1}^{t_2}\!\dd t\,
A_0^e(t,\bm R)T_{\rm adj}^e\right)\right]^{cd},
\quad
\Phi_{\rm adj}\to D(t_2)\Phi_{\rm adj}D^\dagger(t_1).
\label{eq:temporal-adjoint-line}
\end{equation}
Here $A_0$ is the dynamical gauge connection entering the zero-background
static-source transfer problem; it is distinct from the classical
matching background $\bar A_\mu$ introduced above.
Accordingly, in a reflection-positive Euclidean regulator, $\langle{\mathcal O}_A^{c\dagger}(t_2)\Phi_{\rm adj}^{cd}(t_2,t_1){\mathcal O}_B^d(t_1)\rangle/(N_c^2-1)$ has a transfer-matrix spectral representation.  Temporal Wilson lines are propagators of static sources, so a common adjoint self-energy is subtracted in the same matching scheme before energy differences are formed.  With that subtraction, discrete and continuum states satisfy
\begin{align}
\langle A,c|B,d\rangle_{\rm adj}&=\delta^{cd}\delta_{AB},
\notag\\[-0.2em]
\langle\Delta,\lambda,c|\Delta',\lambda',d\rangle_{\rm adj}
&=\delta^{cd}\delta_{\lambda\lambda'}\delta(\Delta-\Delta'),
\label{eq:adjoint-spectral-normalization}
\end{align}
and the corresponding sums and integrals resolve the identity on $\mathcal K_{\rm adj}$.

Singlet projection maps the auxiliary fiber to physical BOEFT states.  If $V_8$ is the heavy-pair octet color fiber, the invariant subspace of $V_8\otimes\mathcal K_{\rm adj}$ is mapped isometrically into the physical BOEFT Hilbert space by
\begin{equation}
{\mathcal J}:\quad
\frac{1}{\sqrt{N_c^2-1}}\sum_c|8,c\rangle\otimes|A,c\rangle
\longmapsto |A\rangle_{\rm BO},
\qquad {\mathcal J}^\dagger{\mathcal J}=1 .
\label{eq:singlet-map}
\end{equation}
The map ${\mathcal J}$ identifies the hybrid-like sector represented
by the adjoint static-source construction.  The extension to the full
complementary sector is a matching identification of the chosen EFT
Hilbert space, rather than a consequence of the adjoint-source
construction alone.  We assume that, after adjoining all
complementary gauge-invariant channels retained in $H_0$---including
open-flavor channels and any additional quarkonium channels included
in the chosen coupled EFT---the auxiliary representation can be
completed to a unitary identification with the physical complementary
sector.

Let $\mathcal H_Q^{\rm aux}$ denote this completed auxiliary space.
Extend ${\mathcal J}$ by the corresponding identities on the
gauge-invariant complementary channels and denote the resulting map
by $\widetilde{\mathcal J}$.  Thus
\[
\widetilde{\mathcal J}^{\dagger}\widetilde{\mathcal J}
 =1_{\mathcal H_Q^{\rm aux}},
\qquad
Q_\phi\widetilde{\mathcal J}=\widetilde{\mathcal J},
\qquad
\operatorname{Ran}\widetilde{\mathcal J}
 =\operatorname{Ran}Q_\phi .
\]
The excitation operator entering the response is then the physical
pullback
\begin{equation}
\Delta_Q\equiv
\widetilde{\mathcal J}^{\dagger}
Q_\phi(H_0-E_\phi)Q_\phi
\widetilde{\mathcal J}
\quad\hbox{on }\mathcal H_Q^{\rm aux}.
\label{eq:physical-pullback}
\end{equation}
It is therefore unitarily equivalent to the restriction of
$H_0-E_\phi$ to the physical complementary sector and is
self-adjoint.

Thus no choice of the Hamiltonian away from the physical
singlet-projected complementary sector enters the response.  For
color-resolved source insertions, the physical pullback has the
canonical color-degenerate lift
\[
\widehat{\mathcal H}_Q^{\rm aux}
 =\mathcal V_{\rm end}\otimes\mathcal H_Q^{\rm aux},
\qquad
\widehat\Delta_Q
 =\bm 1_{\rm end}\otimes\Delta_Q,
\qquad
\mathcal V_{\rm end}\simeq\mathbb C^{N_c^2-1},
\]
where $\mathcal V_{\rm end}$ carries the open adjoint endpoint index
$c$.  This lift is completely fixed by the physical operator
$\Delta_Q$ and introduces no additional dynamics or physical channel.
Whenever explicit adjoint source indices occur in a resolvent matrix
element below, $\Delta_Q$ denotes the lifted operator
$\widehat\Delta_Q$.  The temporal adjoint line transports the open
endpoint index between time slices.  Gauge-invariant open-flavor
channels and their zero-field mixing are retained in
$H_0$~\cite{OpenFlavorBO2026}.

We adapt the normalized source-residue prescription of
Ref.~\cite{ResonancePolarizability2026} to the static-source BO sector.
The Euclidean transfer construction above is used to define the
zero-background spectrum and state normalization.  The causal response
is obtained by analytic continuation in the external energy; the
$i0$ prescriptions below refer to the corresponding Minkowski boundary
values.

For discrete gauge-invariant external states, let
$G_{BA}(E',E;\bar A)$ be the double Fourier transform of the connected
two-time correlator
$\langle0|{\mathcal T}{\mathcal O}_B(t')
{\mathcal O}_A^\dagger(t)|0\rangle_{\bar A}$.
The external-state poles are defined in the zero-background theory,
\[
G_{XX}(E;0)=
\frac{Z_X}{E-E_X+i0}+G_{XX}^{\rm reg}(E),
\qquad X=A,B.
\]
A generic finite time-dependent background need not preserve
time-translation invariance or stationary poles at $E_A$ and $E_B$.
We therefore define pole extraction coefficientwise, after taking
source derivatives at zero background.  For any function
$F_{BA}(E',E)$ with the corresponding external poles, define
\begin{equation}
{\mathfrak P}_{B\leftarrow A}[F]
\equiv
Z_B^{-1/2}
\lim_{\substack{E'\to E_B\\E\to E_A}}
(E'-E_B+i0)\,
F_{BA}(E',E)\,
(E-E_A+i0)
Z_A^{-1/2}.
\label{eq:normalized-residue}
\end{equation}
For independent admissible background directions
$\bar A_1,\ldots,\bar A_n$, the normalized $n$-source coefficient is
\[
{\mathcal R}^{(n)}_{B\leftarrow A}
[\bar A_1,\ldots,\bar A_n]
\equiv
{\mathfrak P}_{B\leftarrow A}\!
\left[
\left.
\frac{\partial^n}
{\partial\eta_1\cdots\partial\eta_n}
G_{BA}\!\left(
E',E;
\sum_{r=1}^n\eta_r\bar A_r
\right)
\right|_{\boldsymbol\eta=0}
\right].
\]
Thus the source derivatives are taken in the zero-background theory
before the normalized external poles are amputated.

For an adjoint endpoint, the outgoing projection is taken in
$\mathcal K_{\rm adj}$ using
Eqs.~\eqref{eq:temporal-adjoint-line}
and~\eqref{eq:adjoint-spectral-normalization}.
A discrete transfer eigenvalue gives the same residue prescription;
a continuum coefficient is the energy-normalized matrix element onto
$|\Delta,\lambda,c\rangle$ with its energy-conserving delta function
removed.  The time integration makes $\mathcal R$ dimensionless.
The partonic cut considered later introduces perturbative-gluon LSZ
factors only at the corresponding partonic projection.

QCD-to-BOEFT matching is imposed coefficientwise in the source
expansion,
\begin{equation}
{\mathcal R}_{B\leftarrow A}^{{\rm QCD},(n)}
=
{\mathcal R}_{B\leftarrow A}^{{\rm BO},(n)}
+O\!\left[(Q_{\rm soft}/\Lambda_{\rm match})^p\right].
\label{eq:background-matching}
\end{equation}
Here the remainder is schematic: $p$ denotes the first omitted order
in the particular multipole, $1/m_Q$, or matching expansion being
considered, rather than a single universal expansion parameter.
For the leading local $E1$ term and the quadratic elastic response, coefficient extraction gives, with $\bar{\mathcal E}(\omega)=\int\dd t\,e^{i\omega t}\bar{\mathcal E}(t)$,
\begin{align}
{\mathcal R}_{A^c\leftarrow\phi}^{(1)}(\omega)
&=-d_{Ai}^{ca}\,\bar{\mathcal E}_i^a(\omega),
\notag\\[-0.2em]
{\mathcal R}_{\phi\leftarrow\phi}^{(2)}
&=-\frac12\int\frac{\dd\omega}{2\pi}\,
\bar{\mathcal E}_i^a(-\omega)\alpha_{{\rm BO},ij}^{ab}(\omega)
\bar{\mathcal E}_j^b(\omega).
\label{eq:background-coefficients}
\end{align}
The linear coefficient $d_{Ai}^{ca}$ and the quadratic kernel
$\alpha_{{\rm BO},ij}^{ab}$ are the $n=1$ and $n=2$ source
coefficients defined by the coefficientwise pole prescription above.
Independent background directions are used, with mixed quadratic
components reconstructed by polarization.  Independent
admissible background directions are used, with mixed quadratic
components reconstructed by polarization.  Background-gauge covariance requires
\begin{equation}
d_{Ai}^{ca}\longrightarrow
D_{\rm adj}^{cc'}d_{Ai}^{c'a'}D_{\rm adj}^{\dagger a'a},
\label{eq:intertwiner}
\end{equation}
so $d^{ca}$ is an adjoint intertwiner.  Its contraction through Eq.~\eqref{eq:singlet-map} and the elastic quadratic term are gauge invariant.  At zero background, global color symmetry makes $d^{ca}$ proportional to $\delta^{ca}$ within each adjoint copy and $\alpha_{{\rm BO},ij}^{ab}$ proportional to $\delta^{ab}$ for the singlet response.  For the fixed external state $|\phi\rangle$, the response coefficient
$\alpha_{{\rm BO},ij}^{ab}$ is a c-number kernel.  Since $[\bar{\mathcal E}(\omega)]=M$, the dimensionless residues give $[d_{Ai}^{ca}]=M^{-1}$ and $[\alpha_{{\rm BO},ij}^{ab}]=M^{-3}$ for discrete normalization; a continuum normalization transfers compensating dimensions between $d$ and its spectral measure.

The same matching first defines covariant transition components on the
preprojection adjoint fiber,
\begin{equation}
\langle A,c|d_i^a\rangle_{\rm cov}\equiv d_{Ai}^{ca}.
\label{eq:matched-dipole-components}
\end{equation}
Let
\[
|A\rangle_Q\equiv\widetilde{\mathcal J}^{\dagger}
|A\rangle_{\rm BO}
\]
denote the corresponding state in $\mathcal H_Q^{\rm aux}$.  For each
external source index $a$, the coefficients in
Eq.~\eqref{eq:matched-dipole-components} define a vector in the lifted
space by
\[
|d_i^a\rangle
 =\sum_{A,c}d_{Ai}^{ca}\,
 |c\rangle_{\rm end}\otimes|A\rangle_Q ,
\]
where the sum over $A$ includes the corresponding spectral integral for
continuum-normalized states.  Equivalently,
\[
\begin{aligned}
\langle d_i^a|(\widehat\Delta_Q-z)^{-1}|d_j^b\rangle
={}&\sum_{A,B,c}
(d_{Ai}^{ca})^*
\langle A|(\Delta_Q-z)^{-1}|B\rangle_Q
d_{Bj}^{cb},
\end{aligned}
\]
with sums replaced by spectral integrals where appropriate.  At zero
background, global color symmetry gives
\[
d_{Ai}^{ca}=\delta^{ca}d_{Ai},
\qquad
|d_i^a\rangle=|a\rangle_{\rm end}\otimes|d_i\rangle,
\qquad
|d_i\rangle=\sum_A d_{Ai}|A\rangle_Q,
\]
and therefore
\[
\alpha_{{\rm BO},ij}^{ab}
 =\delta^{ab}\alpha_{{\rm BO},ij}.
\]

Equations~\eqref{eq:adjoint-inner-product}--\eqref{eq:background-matching} define the source-dependent BOEFT vertex through Euclidean static-source correlators.  Before this fixed singlet contraction, the leading-$E1$ short-distance component is
\begin{equation}
|d_i^a\rangle_{\rm sd}=
\sqrt{\frac{T_F}{N_c}}\,
Q_o\!\left[
|O^a\rangle\otimes
V_A(r;\mu_s)r_i|\phi\rangle_{\rm rel}
\right]
+O(r^2,1/m_Q).
\label{eq:short-distance-dipole}
\end{equation}
Here $|\phi\rangle_{\rm rel}$ is the singlet relative-coordinate wave
function appearing in Eq.~\eqref{eq:octet-response}, while
$|O^a\rangle=T^a/\sqrt{T_F}$ denotes the normalized octet color basis
vector in the standard matrix representation.  The operator $Q_o$
projects onto the pNRQCD octet sector.  Applying the adjoint-fiber lift
described above gives the $|d_i^a\rangle$ used in the resolvent matrix
elements.  Equation~\eqref{eq:matched-dipole-components} supplies the full transition vector for a mixed BO eigenstate, while Eq.~\eqref{eq:short-distance-dipole} is its leading short-distance component.  Higher-time-derivative source vertices enter at subsequent orders.

At fixed quadratic order in the background and for the specified
leading local $E1$ one-field source, insertion of complementary states
gives the connected response at this operator order,
\begin{align}
\alpha_{{\rm BO},ij}^{ab}(\omega)={}&K_{ij}^{ab}(\omega)
+\langle d_i^a|(\Delta_Q-z_+)^{-1}|d_j^b\rangle\notag\\[-0.2em]
&+\langle d_j^b|(\Delta_Q-z_-)^{-1}|d_i^a\rangle,
\qquad z_+=\omega+i0,\quad z_-=-\omega-i0.
\label{eq:matched-response}
\end{align}
The two-source kernel $K_{ij}^{ab}$ collects contributions quadratic
in the background that are irreducible with respect to a cut through
one complementary-sector propagator at the stated operator order.
Corrections that modify a one-field source vertex belong instead to
the matched vectors $|d_i^a\rangle$---equivalently, to the source map
$\mathsf D$ introduced below---and generate reducible terms through the
resolvent.
Genuine local two-field matching operators contribute to $K$, whereas
higher multipoles carrying additional background gradients populate
separate higher-dimension response tensors.  For the purely electric response generated by the leading-$E1$
pNRQCD interaction, this irreducible two-source kernel vanishes at
tree level.  Equation~\eqref{eq:matched-response} is the QCD-matched response to be reduced to the Peskin OPE.

\subsection{Projected resolvent and channel separation}

For BO channels retained explicitly and modes removed at the next matching scale, use the standard Feshbach projection~\cite{Feshbach1958}.  Let $1_Q=Q_R+Q_H$ on $\mathcal H_Q^{\rm aux}$, where $Q_R=Q_R^\dagger$, $Q_H=Q_H^\dagger$, and $Q_RQ_H=0$.  When source-color indices are displayed, $Q_R$, $Q_H$, and the blocks
$\Delta_{XY}$ below denote their corresponding adjoint-fiber lifts,
$\bm 1_{\rm end}\otimes Q_{R,H}$ and
$\bm 1_{\rm end}\otimes\Delta_{XY}$. The off-diagonal Hamiltonian blocks are retained.  In the corresponding block basis,
\begin{equation}
\Delta_Q=\begin{pmatrix}\Delta_{RR}&\Delta_{RH}\\
\Delta_{HR}&\Delta_{HH}\end{pmatrix},\qquad
|d_i^a\rangle=\binom{|d_{Ri}^a\rangle}{|d_{Hi}^a\rangle}.
\label{eq:block-hamiltonian}
\end{equation}
To express the full complementary-sector resolvent in terms of the
retained subspace $Q_R$, introduce the $Q_H$-sector resolvent and the
corresponding energy-dependent Schur complement,
\[
G_H(z)\equiv(\Delta_{HH}-z)^{-1},
\qquad
S_R(z)\equiv
\Delta_{RR}-z-\Delta_{RH}G_H(z)\Delta_{HR}.
\]
Here $G_H$ is the resolvent of the isolated $Q_H$ block, whereas $S_R(z)^{-1}$ is the exact $Q_R$--$Q_R$ block of the full resolvent after
algebraic elimination of $Q_H$.  No low-energy or local expansion has
been made at this stage.  The dressing of the source vectors by the
$Q_H$ sector is incorporated through the associated effective Schur
vertices,
\begin{align}
|d_{Ri}^{a,{\rm eff}}(z)\rangle&=|d_{Ri}^a\rangle-
\Delta_{RH}G_H(z)|d_{Hi}^a\rangle,\notag\\
\langle\widetilde d_{Ri}^{a,{\rm eff}}(z)|&=\langle d_{Ri}^a|-
\langle d_{Hi}^a|G_H(z)\Delta_{HR}.
\label{eq:feshbach-vertices}
\end{align}
The tilde distinguishes the left Schur vertex from the Hermitian adjoint of the right vertex at complex $z$; they coincide in the real resolvent domain.
The standard Schur-complement identity gives
\begin{equation}
\boxed{\begin{aligned}
\langle d_i^a|(\Delta_Q-z)^{-1}|d_j^b\rangle={}&
\langle\widetilde d_{Ri}^{a,{\rm eff}}(z)|S_R(z)^{-1}
|d_{Rj}^{b,{\rm eff}}(z)\rangle\\[-0.2em]
&+\langle d_{Hi}^a|G_H(z)|d_{Hj}^b\rangle .
\end{aligned}}
\label{eq:feshbach-identity}
\end{equation}
This identity retains the off-diagonal mixing
$\Delta_{RH}=Q_R\Delta_QQ_H$ and
$\Delta_{HR}=Q_H\Delta_QQ_R$ in both the retained propagator and its
source residues.  It is initially understood for $z$ in the resolvent
sets of both $\Delta_{HH}$ and $\Delta_Q$; physical pole and cut
boundary values are obtained by the corresponding limiting
continuation.

Applying the first term of Eq.~\eqref{eq:feshbach-identity} to the two
resolvents in Eq.~\eqref{eq:matched-response} defines the resolved
response explicitly as
\[
\begin{aligned}
\alpha_{R,ij}^{ab}(\omega)\equiv{}&
\langle\widetilde d_{Ri}^{a,{\rm eff}}(z_+)|
S_R(z_+)^{-1}
|d_{Rj}^{b,{\rm eff}}(z_+)\rangle\\
&+
\langle\widetilde d_{Rj}^{b,{\rm eff}}(z_-)|
S_R(z_-)^{-1}
|d_{Ri}^{a,{\rm eff}}(z_-)\rangle .
\end{aligned}
\]
The complementary contribution is therefore
\begin{equation}
\begin{aligned}
C_{ij}^{ab}(\omega)
&\equiv
\alpha_{{\rm BO},ij}^{ab}(\omega)
-\alpha_{R,ij}^{ab}(\omega)\\
&=
K_{ij}^{ab}(\omega)
+\langle d_{Hi}^a|G_H(z_+)|d_{Hj}^b\rangle
+\langle d_{Hj}^b|G_H(z_-)|d_{Hi}^a\rangle .
\end{aligned}
\label{eq:contact-subtraction}
\end{equation}
This exact subtraction assigns every contribution once: the
$S_R(z)^{-1}$ terms carry propagation through channels retained
dynamically, whereas $C_{ij}^{ab}$ contains the direct $Q_H$
propagation and the irreducible two-source kernel.  With these
definitions, the exact response is
\begin{equation}
\alpha_{{\rm BO},ij}^{ab}(\omega)
=
\alpha_{R,ij}^{ab}(\omega)
+C_{ij}^{ab}(\omega).
\label{eq:bo-response}
\end{equation}
Equation~\eqref{eq:bo-response} is an exact algebraic decomposition
and does not imply that either contribution has already been expanded
locally. The response topology and its exact channel separation are summarized
in Fig.~\ref{fig:response-topology}.
\begin{figure}[H]
\centering
\resizebox{\linewidth}{!}{%
\begin{tikzpicture}[
  font=\small,
  >=Latex,
  state/.style={draw,rounded corners=2pt,minimum width=9mm,
    minimum height=6mm,fill=gray!8},
  source/.style={circle,draw,thick,minimum size=6mm,inner sep=0pt,
    fill=white},
  resolvent/.style={draw,thick,rounded corners=2pt,minimum width=31mm,
    minimum height=8mm,fill=blue!8},
  kernel/.style={draw,thick,ellipse,minimum width=19mm,
    minimum height=8mm,fill=orange!12},
  local/.style={draw,thick,rounded corners=2pt,minimum width=26mm,
    minimum height=8mm,fill=green!9},
  field/.style={->,thick,teal!65!black},
  prop/.style={->,thick},
  mix/.style={->,thick,dashed,purple!70!black}
]
\node[anchor=west,font=\bfseries] at (0,3.45)
  {(a) Full matched response};

\node[state] (pL) at (0.7,2.15) {$\Phi$};
\node[source] (dJ) at (2.15,2.15) {$D_j^b$};
\node[resolvent] (RQ) at (4.45,2.15)
  {$R_Q(z_+)=(\Delta_Q-z_+)^{-1}$};
\node[source] (dI) at (6.75,2.15) {$D_i^a$};
\node[state] (pR) at (8.20,2.15) {$\Phi$};

\draw[prop] (pL)--(dJ);
\draw[prop] (dJ)--(RQ);
\draw[prop] (RQ)--(dI);
\draw[prop] (dI)--(pR);
\draw[field] (2.15,3.00)--node[right] {$\bar E_j^b$} (dJ);
\draw[field] (6.75,3.00)--node[right] {$\bar E_i^a$} (dI);

\node[align=left,anchor=west] at (8.75,2.15)
  {$+(i,a,z_+)\leftrightarrow(j,b,z_-)$};

\node[state] (kL) at (1.45,0.65) {$\Phi$};
\node[kernel] (K) at (4.45,0.65) {$K_{ij}^{ab}$};
\node[state] (kR) at (7.45,0.65) {$\Phi$};

\draw[prop] (kL)--(K);
\draw[prop] (K)--(kR);
\draw[field] (3.85,1.45)--node[left] {$\bar E_j^b$} (K.north west);
\draw[field] (5.05,1.45)--node[right] {$\bar E_i^a$} (K.north east);

\node at (0.65,0.65) {$+$};
\node[align=left,anchor=west] at (8.75,0.65)
  {irreducible with respect to\\one $Q$-sector resolvent};

\draw[gray!45] (0,-0.15)--(15.7,-0.15);

\node[anchor=west,font=\bfseries] at (0,-0.65)
  {(b) Exact Feshbach separation $Q=Q_R\oplus Q_H$};

\node[source] (deL) at (1.50,-1.65)
  {$\widetilde D_R^{\rm eff}$};
\node[resolvent] (SR) at (4.45,-1.65)
  {$S_R(z)^{-1}$};
\node[source] (deR) at (7.40,-1.65)
  {$D_R^{\rm eff}$};

\draw[prop] (deL)--(SR);
\draw[prop] (SR)--(deR);

\node[anchor=west] at (8.25,-1.65)
  {$\alpha_R$: retained channels remain dynamical};

\node[source] (dhL) at (1.50,-3.00) {$D_H$};
\node[resolvent] (GH) at (4.45,-3.00) {$G_H(z)$};
\node[source] (dhR) at (7.40,-3.00) {$D_H$};

\draw[prop] (dhL)--(GH);
\draw[prop] (GH)--(dhR);

\node[kernel] (Kb) at (10.15,-3.00) {$K$};
\node at (8.50,-3.00) {$+$};
\node[anchor=west] at (11.35,-3.00)
  {$C$: direct $Q_H$ propagation plus kernel};

\draw[mix] (GH.north)
  --node[right] {$\Delta_{RH},\Delta_{HR}$} (SR.south);

\draw[gray!45] (0,-3.80)--(15.7,-3.80);

\node[anchor=west,font=\bfseries] at (0,-4.30)
  {(c) Subsequent reductions};

\node[local] (gap) at (1.65,-5.75)
  {spectrally separated $Q_H$};
\node[local] (mom) at (5.15,-5.75)
  {inverse moments};
\node[local] (ope) at (8.65,-5.75)
  {local operators};

\draw[prop] (gap)--(mom);
\draw[prop] (mom)--(ope);

\node[
  font=\footnotesize,
  fill=white,
  inner sep=1pt
] at (3.40,-5.10)
  {$G_H(z)$ expansion};

\node[resolvent] (wc) at (1.65,-7.10)
  {weak-coupling sector};
\node[resolvent] (oct) at (5.15,-7.10)
  {octet resolvent};
\node[local] (bp) at (8.65,-7.10)
  {BP moments and cut};

\draw[prop] (wc)--(oct);
\draw[prop] (oct)--(bp);

\node[align=left,anchor=west] at (10.35,-6.42)
  {$Q_R$ is not expanded locally;\\
   the BP limit requires the additional\\
   weak-coupling/free-octet assumptions.};
\end{tikzpicture}%
}
\caption{Operator-topology representation of the matched BOEFT
response.  (a) Two one-field source insertions connected by the full
complementary-sector resolvent, together with the crossed contribution
and the irreducible two-field kernel $K$.  (b) Exact Feshbach
decomposition $Q=Q_R\oplus Q_H$: the retained channels propagate
through $S_R^{-1}$ with $Q_H$-dressed source vertices, whereas direct
$Q_H$ propagation and $K$ form the complementary response $C$.
(c) (c) Only after spectral separation of the eliminated $Q_H$ block may
its $G_H$ dependence be expanded into local matching corrections and
inverse moments.  The lines denote full
resolvents and source maps, not perturbative free-particle
propagators.}
\label{fig:response-topology}
\end{figure}

Integrating out $Q_H$ requires a spectral separation between the
expansion domain and $\sigma(\Delta_{HH})$.  For an expansion about
$z=0$, a sufficient condition is
$|z|\,\|\Delta_{HH}^{-1}\|<1$, under which
\[
G_H(z)
=
\Delta_{HH}^{-1}
\sum_{n=0}^{\infty}
\left(z\Delta_{HH}^{-1}\right)^n .
\]
This expansion must be made in every occurrence of $G_H$: in the
Schur complement $S_R$, in both effective source vertices, and in the
direct $Q_H$ term in Eq.~\eqref{eq:contact-subtraction}.  If
$K_{ij}^{ab}(z)$ is analytic in the same domain, it is expanded there
as well.  Eliminating $Q_H$ therefore generates local corrections to
the retained propagation and source vertices inside $\alpha_R$, as
well as to the contact terms collected in $C$.

After these $Q_H$-dependent functions have been expanded, the final
inverse $S_R(z)^{-1}$ is left unexpanded and retains the poles and cuts
of the channels kept dynamically.  By contrast, the complete
remainder $C_{ij}^{ab}$ admits an expansion in local contact
coefficients when the eliminated-sector resolvent and
$K_{ij}^{ab}$ are analytic in the matching domain.  The full
inverse-moment expansion developed below is the stronger limit in
which no source-accessible singularity is retained explicitly and the
complete source-generated spectrum is separated from the expansion
point.

\subsection{Spectral moments and the local operator basis}

Independently of whether the full response admits a local expansion,
the self-adjoint excitation operator $\Delta_Q$ has the spectral
resolution
\[
\begin{aligned}
1_Q&=\int_{\sigma(\Delta_Q)}
       \dd P_{\Delta_Q}(\Delta),\\
\Delta_Q&=\int_{\sigma(\Delta_Q)}
       \Delta\,\dd P_{\Delta_Q}(\Delta),\\
(\Delta_Q-z)^{-1}
&=\int_{\sigma(\Delta_Q)}
  \frac{\dd P_{\Delta_Q}(\Delta)}{\Delta-z}.
\end{aligned}
\]
Here $\Delta$ is the real spectral variable of $\Delta_Q$, with the
meaning of a complementary-sector excitation energy; it is not a
separate operator.  The projector-valued spectral measure
$\dd P_{\Delta_Q}(\Delta)$ defines the source-weighted,
matrix-valued dipole measure
\begin{equation}
\dd\mu_{ij}^{ab}(\Delta)=
\langle d_i^a|\dd P_{\Delta_Q}(\Delta)|d_j^b\rangle
=\delta^{ab}\dd\mu_{ij}(\Delta),
\label{eq:measure}
\end{equation}
where the last equality follows for a color-singlet external state.
The measure is positive semidefinite in the matrix sense:
$u_{ia}^*\dd\mu_{ij}^{ab}u_{jb}\geq0$ for every complex source
tensor $u_{ia}$; an off-diagonal component need not be positive
separately.  It combines the density of complementary-sector states
with their QCD-matched transition strengths, supplementing the
information contained in the BO potentials.

Global color invariance likewise gives
$\alpha_{{\rm BO},ij}^{ab}=\delta^{ab}\alpha_{{\rm BO},ij}$ and
$K_{ij}^{ab}=\delta^{ab}K_{ij}$.  Before any local expansion, the exact color-singlet response admits
the analytic spectral representation
\begin{equation}
\alpha_{{\rm BO},ij}(z)=
\int_{\sigma(\Delta_Q)}\!\left[
\frac{\dd\mu_{ij}(\Delta)}{\Delta-z}
+\frac{\dd\mu_{ji}(\Delta)}{\Delta+z}
\right]
+K_{ij}(z),
\qquad \operatorname{Im}z\neq0 .
\label{eq:spectral}
\end{equation}
Before the color-singlet projection, the first and second spectral
terms contain $\dd\mu_{ij}^{ab}$ and $\dd\mu_{ji}^{ba}$,
respectively, and the irreducible kernel is $K_{ij}^{ab}$.

The physical upper-boundary response appearing in
Eq.~\eqref{eq:matched-response} is
\[
\alpha_{{\rm BO},+,ij}(\omega)
\equiv
\lim_{\epsilon\downarrow0}
\alpha_{{\rm BO},ij}(\omega+i\epsilon),
\]
and analogously
$\alpha_{{\rm BO},-,ij}(\omega)$ is obtained from the lower boundary.
Define the full distributional spectral density
\[
\varrho_{ij}(\omega)\equiv
\int_{\sigma(\Delta_Q)}
\delta(\omega-\Delta)\,\dd\mu_{ij}(\Delta).
\]
It includes both the absolutely continuous and pure-point parts of
the source-weighted spectral measure.  The Sokhotski--Plemelj formula
then gives
\begin{equation}
\begin{aligned}
\operatorname{Disc}\alpha_{{\rm BO},ij}(\omega)
&\equiv
\alpha_{{\rm BO},+,ij}(\omega)
-\alpha_{{\rm BO},-,ij}(\omega)\\
&=
2\pi i\left[
\varrho_{ij}(\omega)-\varrho_{ji}(-\omega)
\right]
+\operatorname{Disc}K_{ij}(\omega),
\end{aligned}
\label{eq:boundary-jump}
\end{equation}
where
\[
\operatorname{Disc}K_{ij}(\omega)
\equiv K_{+,ij}(\omega)-K_{-,ij}(\omega).
\]
On an absolutely continuous component,
$\varrho_{ij}(\omega)=\rho_{ij}(\omega)$.  An isolated eigenvalue
$\Delta_n$ contributes
\[
\varrho_{ij}(\omega)\supset
w_{ij}^{(n)}\delta(\omega-\Delta_n),
\qquad
\delta^{ab}w_{ij}^{(n)}
\equiv
\langle d_i^a|
P_{\Delta_Q}(\{\Delta_n\})
|d_j^b\rangle .
\]

Thus Eq.~\eqref{eq:boundary-jump} contains both the continuum-cut
discontinuity and the discrete-pole contributions, including their
crossed negative-frequency counterparts.  Equation~\eqref{eq:spectral}
is therefore not itself a local approximation.

To determine the spectral radius available for a local expansion,
introduce the positive scalar trace measure
\[
\dd\mu_{\rm tr}(\Delta)
\equiv
\sum_i\dd\mu_{ii}(\Delta).
\]
Because $\dd\mu_{ij}$ is positive semidefinite, the trace measure has
the same support as the full source-weighted matrix measure.  Its
support consists of the excitation energies for which every
neighborhood carries nonzero total dipole-source strength.  Define
\begin{equation}
\Delta_{\rm src}\equiv
\operatorname{dist}\!\left(
0,\operatorname{supp}\mu_{\rm tr}
\right).
\label{eq:source-gap}
\end{equation}
For the stable ground-state situation relevant to the Peskin
reduction,
\[
\operatorname{supp}\mu_{\rm tr}
\subseteq[\Delta_{\rm src},\infty),
\qquad
\Delta_{\rm src}>0,
\]
so that
$\Delta_{\rm src}=\inf\operatorname{supp}\mu_{\rm tr}$ is the lowest
excitation energy carrying nonzero dipole-source strength.

For $\omega>0$ in this positive-spectrum situation,
$\varrho_{ji}(-\omega)=0$.  On an absolutely continuous component,
write $\varrho_{ij}=\rho_{ij}$.  For a time-reversal-invariant
isotropic state,
\[
\rho_{ij}(\omega)
=\frac{\delta_{ij}}{3}\rho_{\rm tr}(\omega),
\qquad
\rho_{\rm tr}(\omega)\geq0,
\]
and, away from isolated poles,
\[
\operatorname{Im}\alpha_{{\rm BO},+,ij}(\omega)
=
\delta_{ij}\frac{\pi\rho_{\rm tr}(\omega)}{3}
+\operatorname{Im}K_{+,ij}(\omega).
\]

The fully local reduction makes the additional matching choice that
no source-accessible pole or cut is retained as an explicit
low-energy degree of freedom.  If a low-lying singularity must remain
dynamical, one instead returns to the Feshbach decomposition
Eq.~\eqref{eq:bo-response}.  The eliminated-sector resolvent $G_H(z)$
is then expanded wherever it occurs: in the Schur complement, in the
effective source vertices, and in the direct $Q_H$ contribution,
while the final retained-sector inverse $S_R(z)^{-1}$ is left
unexpanded.  Thus elimination of $Q_H$ produces local corrections to
the retained Hamiltonian and source vertices as well as local contact
terms in $C$.  The analytic part of $K_{ij}(z)$ is expanded in the
same matching domain.  A fully local expansion of the complete
response additionally requires $K_{ij}(z)$ to be analytic for
$|z|<\Delta_{\rm src}$.

Under these assumptions, write the Taylor expansion of the
irreducible kernel as
\[
K_{ij}(z)=\sum_{m=0}^{\infty}\kappa_{ij}^{(m)}z^m,
\qquad
\kappa_{ij}^{(m)}
=\frac{1}{m!}
\left.\frac{\dd^mK_{ij}}{\dd z^m}\right|_{z=0}.
\]
Expansion of Eq.~\eqref{eq:spectral} for
$|\omega|<\Delta_{\rm src}$ then gives the fully local response
\begin{align}
\alpha_{{\rm BO},ij}(\omega)={}&
2\sum_{k=0}^{\infty}\omega^{2k}
\int\frac{\dd\mu_{(ij)}(\Delta)}{\Delta^{2k+1}}
\notag\\
&+2\sum_{k=0}^{\infty}\omega^{2k+1}
\int\frac{\dd\mu_{[ij]}(\Delta)}{\Delta^{2k+2}}
+\sum_{m=0}^{\infty}\kappa_{ij}^{(m)}\omega^m .
\label{eq:moments}
\end{align} 
Parentheses and brackets denote the symmetric and antisymmetric parts
in the spatial indices.  Time-reversal and crossing symmetry require
the irreducible kernel to satisfy
\[
K_{ij}(z)=K_{ji}(-z).
\]
Consequently, for its rotational-scalar projection,
\[
K_{\rm sc}(z)\equiv\frac{1}{3}\delta_{ij}K_{ij}(z),
\qquad
\kappa_{\rm sc}^{(m)}
 \equiv\frac{1}{3}\delta_{ij}\kappa_{ij}^{(m)},
\]
one has
\[
K_{\rm sc}(z)=K_{\rm sc}(-z),
\qquad
\kappa_{\rm sc}^{(2k+1)}=0.
\]
For a time-reversal-invariant isotropic $S$ wave one also has
$\dd\mu_{[ij]}=0$.  Hence the complete rotational-scalar response,
including both the spectral contribution and the irreducible kernel,
reduces to a crossing-even tower in $\omega^{2k}$.  The operator identities below apply to the scalar
contraction
$E_i^a(iD_0)_{ab}^{n-2}E_i^b$; a general uncontracted $E_iE_j$
response also contains irreducible rotational tensors.

At the level of the displayed two-field tower, the frequency powers
are promoted to symmetrized gauge-covariant derivatives through
$\omega\to iv\!\cdot\!D$.  Commutators of covariant derivatives
generate additional local operators, represented by the ellipsis in
Eq.~\eqref{eq:local-lagrangian}.  The leading-twist projection uses
symmetric-traceless gluon tensors~\cite{Arleo2001}, whereas the
untraced rest-frame electric operator also contains correlated
metric-trace sectors.  Introduce the symmetrized rank-$n$ tensor
\begin{equation}
{\mathcal X}_n^{\mu_1\ldots\mu_n}=
-{\mathcal S}\!\left\{
F^{a\mu_1\alpha}
\bigl[(iD^{\mu_2})\cdots(iD^{\mu_{n-1}})\bigr]_{ab}
F^{b\mu_n}{}_{\alpha}\right\},
\qquad n=2,4,\ldots ,
\label{eq:untraced-tensor}
\end{equation}
where ${\mathcal S}$ symmetrizes the Lorentz indices without
subtracting traces.  At the level of the displayed four-dimensional
two-field operator, the harmonic decomposition
of Ref.~\cite{GeorgiPolitzer1976} reads
\begin{align}
{\mathcal X}_n^{\mu_1\ldots\mu_n}
={}&{\mathcal X}_{n,0}^{\mu_1\ldots\mu_n}
+\sum_{k=1}^{n/2}g^{(\mu_1\mu_2}\cdots
g^{\mu_{2k-1}\mu_{2k}}
{\mathcal X}_{n,k}^{\mu_{2k+1}\ldots\mu_n)},
\notag\\[-0.2em]
{\mathcal Q}_n^E(v)\equiv{}&v_{\mu_1}\cdots v_{\mu_n}
{\mathcal X}_n^{\mu_1\ldots\mu_n}
={\mathcal O}_{n,0}^g(v)+\sum_{k=1}^{n/2}{\mathcal O}_{n,k}^{g,{\rm tr}}(v).
\label{eq:trace-basis}
\end{align}
Each ${\mathcal X}_{n,k}$ is symmetric traceless of rank $n-2k$; all combinatorial coefficients are fixed by the first equality and included in its definition.  The $k=0$ member
\begin{equation}
{\mathcal O}_{n,0}^g(v)=
-v_{\mu_1}\cdots v_{\mu_n}
\left[F^{a\mu_1\alpha}
\bigl[(iD^{\mu_2})\cdots(iD^{\mu_{n-1}})\bigr]_{ab}
F^{b\mu_n}{}_{\alpha}\right]_{\rm ST}
\label{eq:twist-two-operator}
\end{equation}
is the gluon twist-two operator, while $k\geq1$ are the metric-trace components.  We use $g_{\mu\nu}=\operatorname{diag}(1,-1,-1,-1)$, $v=(1,\bm0)$, and $E_i^a=F^{a0i}$, so ${\mathcal Q}_n^E=E_i^a(iD_0)_{ab}^{n-2}E_i^b$.  Already at $n=2$,
\begin{equation}
{\mathcal O}_{2,0}^g=E_a^2+\frac14F_a^2
=\frac12(E_a^2+B_a^2),\qquad
{\mathcal O}_{2,1}^{g,{\rm tr}}=-\frac14F_a^2
=\frac12(E_a^2-B_a^2),
\label{eq:n2-trace-split}
\end{equation}
and their sum is $E_a^2$.  The scalar $F_a^2$ term is an independent
trace-sector operator.  Equations~\eqref{eq:trace-basis}--%
\eqref{eq:n2-trace-split} are strict four-dimensional algebraic
identities and specify the tensor decomposition used for the leading
matching coefficients displayed here.  In a loop calculation
regulated in $d=4-2\epsilon$, the trace projectors and operator basis
must instead be defined in $d$ dimensions before renormalization;
evanescent structures may then contribute finite terms after operator
mixing.  No such $d$-dimensional loop renormalization of the present
tower is performed here.

For $n>2$, the metric traces belong to higher-twist sectors.  Their
renormalization generally requires an enlarged operator basis;
Ref.~\cite{Kodaira1997} illustrates this structure in the
flavor-singlet twist-three $g_2$ problem, including
equation-of-motion and non-gauge-invariant (BRST) operators.
Consequently, the correlated coefficients produced by the
four-dimensional leading matching do not imply common
renormalization-group evolution of the twist-two and metric-trace
sectors.  At $n=2$, the singlet quark and gluon spin-two operators
mix, while the scalar sector is spanned by $F_a^2$ and
$\sum_qm_q\bar q q$ in a specified renormalization
scheme~\cite{Tanaka2019,Panagopoulos2021,ScalarSpinTwo2026}.
The inverse moments in Eq.~\eqref{eq:moments} determine the
resolvent-generated matching contributions in this basis, while the
Taylor coefficients $\kappa_{ij}^{(m)}$ supply additional local
matching contributions from the irreducible kernel.

To fix the Wilson-coefficient normalization, let $\Phi$ denote the
nonrelativistic quarkonium field and use unrescaled chromoelectric and
chromomagnetic fields.  We adopt the standard chromoelectric
normalization of Refs.~\cite{Peskin1979,Luke1992,LakhinaSwanson2004}
and define the chromomagnetic coefficient by the analogous convention:
\begin{equation}
\begin{aligned}
\delta{\mathcal L}_{\rm loc}={}&+a_0^3\Phi^\dagger\Phi
\sum_{N\geq1}\epsilon_0^{\,2-2N}
\left[c_E^{(N)ij}{\mathcal Q}_{2N}^{E,ij}
+c_B^{(N)ij}{\mathcal Q}_{2N}^{B,ij}+\cdots\right],\\[-0.2em]
{\mathcal Q}_{2N}^{E,ij}={}&E_i^a(iD_0)_{ab}^{2N-2}E_j^b,\qquad
{\mathcal Q}_{2N}^{B,ij}=B_i^a(iD_0)_{ab}^{2N-2}B_j^b .
\end{aligned}
\label{eq:local-lagrangian}
\end{equation}
At dimension four, define the physical polarizabilities by
\begin{equation}
\delta{\mathcal L}_{\rm pol}=\frac12\Phi^\dagger\Phi
\left(\alpha_E^{\Phi,ij}E_i^aE_j^a
+\alpha_B^{\Phi,ij}B_i^aB_j^a\right),
\qquad
\alpha_{E,B}^{\Phi,ij}=2a_0^3c_{E,B}^{(1)ij}.
\label{eq:physical-polarizabilities}
\end{equation}
Spatial indices are raised and lowered with $\delta_{ij}$.  For an
isotropic $S$ wave, we write
\[
c_{E,B}^{(N)ij}=\delta^{ij}c_{E,B}^{(N)},
\qquad
\alpha_{E,B}^{\Phi,ij}
=\delta^{ij}\alpha_{E,B}^{\Phi}.
\]
Equivalently, if $\mathcal E=gE$ and $\mathcal B=gB$, the coefficients of $\mathcal E_i^a\mathcal E_j^a/2$ and $\mathcal B_i^a\mathcal B_j^a/2$ are $\beta_{E,B}^{ij}=\alpha_{E,B}^{\Phi,ij}/g^2=a_0^3c_{E,B}^{(1)ij}/(2\pi\alpha_s)$.  Hence the color-diagonal static response $\alpha_{{\rm BO},ij}(0)$ equals $\beta_E^{ij}$ in the convention of Eq.~\eqref{eq:measure}.  The minus sign in the normalized residue, Eq.~\eqref{eq:background-coefficients}, is $-{\int}\delta{\mathcal L}_{\rm pol}$; the static level shift is likewise $\delta E=-\delta{\mathcal L}_{\rm pol}$.  This fixes the sign, factor of two, and coupling convention used below;
for the chromoelectric part it agrees with Ref.~\cite{gWEFT2016}, while the chromomagnetic part is defined analogously.
In this convention Peskin's leading-order even-derivative coefficients, for which the
electric singlet--octet matching coefficient is set to $V_A=1$, are denoted
$c_{E,{\rm BP}}^{(N)ij}$ and read~\cite{Peskin1979,BhanotPeskin1979,Arleo2001,LakhinaSwanson2004}
\begin{equation}
c_{E,{\rm BP}}^{(N)ij}=
\frac{2\pi\alpha_s\epsilon_0^{\,2N-2}}{N_ca_0^3}
\left\langle\phi\left|r^i
\frac{1}{(H_a-E_\phi)^{2N-1}}r^j\right|\phi\right\rangle .
\label{eq:peskin-moments}
\end{equation}
Here $\Delta_a\equiv H_a-E_\phi$ is defined with the same heavy-pair rest-mass subtraction as $E_\phi$; in weakly coupled pNRQCD $H_a=h_o$.  In the strict Bhanot--Peskin convention used below, $a_0=4/(m_QN_c\alpha_s)$ and $\epsilon_0=1/(m_Qa_0^2)$.  The coefficient $c_{E,{\rm BP}}^{(N)}$ multiplies the untraced electric operator ${\mathcal Q}_{2N}^E$ in Eq.~\eqref{eq:trace-basis}.  Relative to Eqs.~\eqref{eq:pnrqcd}--\eqref{eq:octet-response}, where $\mathcal E_i^a=gE_i^a$, the conversion of the quadratic residue $-\alpha_{\mathcal E}^{ij}\mathcal E_i^a\mathcal E_j^a/2$ supplies the factor $T_Fg^2=2\pi\alpha_s$ displayed in Eq.~\eqref{eq:peskin-moments}; retaining electric matching inserts the two ordered $V_A$ factors shown explicitly in Eq.~\eqref{eq:matched-electric-coefficient}.  The label $N=1,2,\ldots$ corresponds to $n=2N$ and to ${\mathcal Q}_{2N}^E$.  At the leading matching order considered here,
Eq.~\eqref{eq:peskin-moments} fixes the coefficient of the untraced
electric operator.  Its strict four-dimensional decomposition through
Eq.~\eqref{eq:trace-basis} therefore produces correlated twist-two and
metric-trace components.  Subsequent loop renormalization must be
performed in the appropriate dimensionally regulated enlarged
operator basis and need not preserve this common coefficient.  In the specific leading forward on-shell free-gluon projection used
to compare with the conventional Bhanot--Peskin moments, the
metric-trace matrix elements vanish.  This is a property of that
projection, not an operator identity; the trace-sector operators
remain present in general hadronic matrix elements.

These operator identities specify the local basis once the expansion
exists.  The dynamical hierarchies required to reduce the general
BOEFT response to this local, channel-factorized form are collected
next.

\subsection{Conditions for the Peskin reduction}

The stronger reduction of the full source response to a local OPE
first requires multipole validity and source-resolvent locality.  Denoting the characteristic external energy or momentum by $Q_{\rm soft}$, these conditions are
\begin{align}
r_\phi Q_{\rm soft}&\ll1,\notag\\
|\omega|,\ Q_{\rm soft}&\ll\Delta_{\rm src}.
\label{eq:local-conditions}
\end{align}
The second condition controls the singularities generated by the
source-resolvent part of the response.  Locality of the complete
response additionally requires the irreducible kernel $K_{ij}(z)$ to
be analytic throughout the same matching domain, as assumed below
Eq.~\eqref{eq:source-gap}.
The additional requirement for a single-reference, channel-factorized
local OPE is light-channel factorization in both propagation and source
creation.  Let $\mathcal C$ denote the source-cyclic subspace defined precisely
below, and assume that it admits a fixed identification
\[
\mathcal C\simeq
\mathcal H_{\rm ref}\otimes\mathcal H_{\rm ch}.
\]
Let $H_{\rm ref}$ denote the reference excitation operator, including
the subtraction of $E_\phi$, and define
\[
\Delta_0\equiv
H_{\rm ref}\otimes\bm 1_{\rm ch},
\qquad
\delta\mathbb H_{\rm light}\equiv\Delta_Q-\Delta_0,
\qquad
R_0(z)\equiv(\Delta_0-z)^{-1}.
\]
Thus $\delta\mathbb H_{\rm light}$ contains the light-channel
splittings, channel mixing, and any BO-label dependence absent from the
single-reference operator.  All these operators are understood with
the adjoint-fiber lift when explicit source-color indices are retained.

Let $\mathcal U$ be the finite-dimensional space of source tensors
$u_i^a$, and regard the matched vertices as the map
\[
\mathsf D:\mathcal U\longrightarrow
\widehat{\mathcal H}_Q^{\rm aux},
\qquad
\mathsf D u\equiv u_i^a|d_i^a\rangle
=(\mathsf D_0+\delta\mathsf D)u.
\]
Here $\mathsf D_0$ is the universal short-distance $E1$ map in
Eq.~\eqref{eq:short-distance-dipole}, including its multiplicative
Wilson coefficient.  Define the leading-source subspace
$\mathcal U_0\equiv(\ker\mathsf D_0)^\perp$.  
For the single-reference factorization considered here, propagation
factorization must be accompanied by source factorization.  Under the
fixed identification
$\mathcal C\simeq\mathcal H_{\rm ref}\otimes\mathcal H_{\rm ch}$,
and suppressing the already understood adjoint-endpoint lift, we
therefore assume that the leading source has a channel-universal form:
there exist a map
$\mathsf D_{\rm ref}:\mathcal U_0\to\mathcal H_{\rm ref}$
and a fixed normalized channel vector
$|\chi_0\rangle\in\mathcal H_{\rm ch}$ such that
\begin{equation}
\mathsf D_0u
=
(\mathsf D_{\rm ref}u)\otimes|\chi_0\rangle,
\qquad
\langle\chi_0|\chi_0\rangle=1,
\qquad u\in\mathcal U_0 .
\label{eq:leading-source-factorization}
\end{equation}
Since
$R_0(z)=(H_{\rm ref}-z)^{-1}\otimes\bm 1_{\rm ch}$, the corresponding
reference response factorizes explicitly,
\begin{equation}
\langle d_0(u)|R_0(z)|d_0(v)\rangle
=
\langle\mathsf D_{\rm ref}u|
(H_{\rm ref}-z)^{-1}
|\mathsf D_{\rm ref}v\rangle .
\label{eq:reference-response-factorization}
\end{equation}
Equations~\eqref{eq:factorization-bound}--%
\eqref{eq:factorization-error} therefore bound the deviation of the
full matched response from an explicitly channel-factorized reference
response.  If more than one independent channel vector already occurs
at leading order, those directions must instead be included in the
leading reference source structure and must not be counted as a small
$\delta\mathsf D$ correction.
A sufficient criterion on
this subspace is
\begin{equation}
\begin{gathered}
q_H\equiv\sup_{z\in\mathcal D}
\|\delta\mathbb H_{\rm light}R_0(z)\|<1,
\\[-0.1em]
q_d\equiv
\sup_{\substack{u\in\mathcal U_0\\u\neq0}}
\frac{\|\delta\mathsf D u\|}
     {\|\mathsf D_0u\|},
\qquad
q_H,q_d\ll1 .
\end{gathered}
\label{eq:factorization-bound}
\end{equation}
The estimates below initially control the leading-source subspace
$\mathcal U_0$.  To extend them to the full source space $\mathcal U$,
one must additionally require
\[
\delta\mathsf D\,\ker\mathsf D_0=0
\]
at the retained order.  Otherwise the correction creates genuinely new
source channels; these directions must be promoted to the leading
source map and $\mathcal U_0$ redefined accordingly.
All operator norms are restricted to $\mathcal C$, defined as the smallest closed subspace containing
\[
\operatorname{Ran}
 \bigl(\mathsf D_0|_{\mathcal U_0}\bigr)
+
\operatorname{Ran}
 \bigl(\delta\mathsf D|_{\mathcal U_0}\bigr)
\] and invariant under $R_0(z)$ and
$\delta\mathbb H_{\rm light}$ for $z\in\mathcal D$.  The matching
domain is assumed to lie a nonzero distance from both spectra,
\[
\operatorname{dist}\!\left(
\mathcal D,\,
\sigma(\Delta_0)\cup\sigma(\Delta_Q)
\right)>0,
\]
so that $R_0(z)$ and $R(z)$ are uniformly bounded on $\mathcal D$.  With $R=(\Delta_Q-z)^{-1}$, the standard Neumann-series resolvent estimate~\cite{Kato1995} gives
\begin{equation}
\begin{aligned}
\|R(z)-R_0(z)\|
&\leq
\|R_0(z)\|\frac{q_H}{1-q_H},
\\
\big|
\langle d(u)|R(z)|d(v)\rangle
-\langle d_0(u)|R_0(z)|d_0(v)\rangle
\big|
&\leq
\|d_0(u)\|\,\|d_0(v)\|\,\|R_0(z)\|
\\[-0.2em]
&\quad\times
\left[
\frac{(1+q_d)^2q_H}{1-q_H}
+2q_d+q_d^2
\right].
\end{aligned}
\label{eq:factorization-error}
\end{equation}
Here $z\in\mathcal D$, $u,v\in\mathcal U_0$,
$|d(u)\rangle=\mathsf D u$, and
$|d_0(u)\rangle=\mathsf D_0u$.  Taking $u$ and $v$ to be source-basis vectors in $\mathcal U_0$
directly controls every response component supported on the
leading-source subspace; the diagonal estimate is recovered by setting
$u=v$.  Components outside $\mathcal U_0$ are covered only after the
extension described above.  For unbounded channel Hamiltonians,
Eqs.~\eqref{eq:factorization-bound}--%
\eqref{eq:factorization-error} are first understood on a common
regulated source-cyclic subspace, for example after finite-volume,
ultraviolet, and finite-energy projection.  Removal of the regulator
preserves the uniform operator-norm estimate only if the regulated
Hamiltonians converge in norm resolvent on the matching domain, the
regulated source maps converge in norm, and a regulator-independent
spectral separation from $\mathcal D$ is maintained.  These
convergence properties are sufficient assumptions; they are not
claimed here to follow automatically for a general BOEFT
Hamiltonian.  If only strong-resolvent convergence is available, then
for source vectors converging in norm one obtains pointwise
convergence of the corresponding matrix elements, not a uniform
operator-norm estimate.  Leading BO-label dependence
\cite{LakhinaSwanson2004} is represented either by retaining the
corresponding channel or by introducing channel-labelled local
operators.

These estimates are sufficient operator-theoretic criteria; the
inequalities themselves do not establish that $q_H$ and $q_d$ are
parametrically small in QCD.  Their smallness must follow from the EFT
hierarchy appropriate to the matching problem.  Define the reference
spectral separation on the matching domain by
\[
\Delta_{\rm ref}^{-1}
\equiv
\sup_{z\in\mathcal D}\|R_0(z)\|.
\]
If $\delta\mathbb H_{\rm light}$ is characterized on the relevant
cyclic subspace by a light scale $\Lambda_{\rm light}$, and the leading
source correction is generated by the first omitted multipole, then
schematically,
\begin{equation}
q_H\sim\frac{\Lambda_{\rm light}}{\Delta_{\rm ref}},
\qquad
q_d\sim r_\phi\Lambda_{\rm light}.
\label{eq:factorization-power-counting}
\end{equation}
Equation
\eqref{eq:factorization-power-counting} is only a power-counting
estimate; Eqs.~\eqref{eq:factorization-bound}--\eqref{eq:factorization-error}
are the precise sufficient bounds.

Together with Eq.~\eqref{eq:local-conditions} and the analyticity of
$K_{ij}$, Eqs.~\eqref{eq:factorization-bound}--
\eqref{eq:factorization-error} control a channel-factorized local OPE.
They do not, however, determine the reference propagator or make its
Wilson coefficients perturbatively calculable.  Moreover, these
bounds constrain only the resolvent-generated part of the response:
the Taylor coefficients $\kappa_{ij}^{(m)}$ of the irreducible kernel
remain independent local matching contributions.  Identification of the Wilson coefficients with the pure Peskin inverse
moments therefore requires $K_{ij}$ to vanish at the retained order, as
it does for the leading electric $E1$ response at tree level.  If
$K_{ij}$ is nonzero but analytic, the local OPE still exists, but its
Wilson coefficients receive the additional contributions
$\kappa_{ij}^{(m)}$ and are no longer given by the pure Peskin moments.

The specific weak-coupling pNRQCD coefficients additionally require
\begin{equation}
m_Qv\sim r_\phi^{-1}\gg\Lambda_{\rm QCD},
\qquad \alpha_s(m_Qv)\ll1,
\label{eq:weak-coupling-conditions}
\end{equation}
so that the compact singlet wave function, the octet Green function, and the short-distance source vertex admit weak-coupling matching.  The long-distance gluonic background may remain nonperturbative; the multipole and response expansions organize its coupling to the compact state.  Within the channel-factorized regime controlled by
Eqs.~\eqref{eq:factorization-bound}--\eqref{eq:factorization-error}
and power counted in Eq.~\eqref{eq:factorization-power-counting},
Eq.~\eqref{eq:weak-coupling-conditions} permits the reference
excitation operator and source map to be chosen in their weak-coupling
pNRQCD forms,
\[
H_{\rm ref}\longrightarrow\Delta_o=h_o-E_\phi,
\qquad
\mathsf D_0\longrightarrow\mathsf D_{E1},
\]
where $\mathsf D_{E1}$ is the leading short-distance source defined in
Eq.~\eqref{eq:short-distance-dipole}.  The strict local Bhanot--Peskin limit used below also takes $\Delta_{\rm src}\sim m_Qv^2\gg Q_{\rm soft}$; when the background varies on a hadronic scale, this is the additional hierarchy $m_Qv^2\gg\Lambda_{\rm QCD}$.  These assumptions produce a weak-coupling pNRQCD local OPE, but not yet
the original Bhanot--Peskin coefficients.  The latter additionally use
the leading electric vertex $V_A=1$ and the large-$N_c$/free-octet
approximation
\[
C_F\longrightarrow\frac{N_c}{2},
\qquad
V_o(r)\longrightarrow0.
\]
Retaining the repulsive octet potential gives the finite-$N_c$ pNRQCD
generalization rather than the strict Bhanot--Peskin moments derived
below.

The scale and propagation--source factorization conditions entering
the strict local reduction may therefore be summarized as
\begin{equation}
\boxed{
r_\phi Q_{\rm soft}\ll1,\qquad
|\omega|,\ Q_{\rm soft}\ll
\Delta_{\rm src}\sim m_Qv^2,\qquad
q_H,q_d\ll1,\qquad
m_Qv\gg\Lambda_{\rm QCD},\qquad
\alpha_s(m_Qv)\ll1 .
}
\label{eq:hierarchy-summary}
\end{equation}
The first condition controls the multipole expansion, the second the
full source-resolvent expansion, the third provides the sufficient propagation--vertex criterion for
channel factorization, and
the last two the perturbative evaluation of the compact singlet,
octet propagator, and short-distance source.

For a background varying on a hadronic scale,
$Q_{\rm soft}\sim\Lambda_{\rm QCD}$, the source-gap condition includes
$m_Qv^2\gg\Lambda_{\rm QCD}$.  Equality with the original
Bhanot--Peskin coefficients further assumes that the irreducible
kernel vanishes at the retained order and that the leading-$E1$,
free-octet approximation specified above is made.  If either condition
is relaxed, the result remains a local pNRQCD OPE, but its coefficients
are generalized matching coefficients rather than the strict
Bhanot--Peskin moments.

The matching domain $\mathcal D$ lies below the physical cut, at a finite distance from the spectrum.  Inverse moments obtained there become local Wilson coefficients for a sector removed in the corresponding matching step.  The absorptive dissociation kernel follows from the discontinuity of the full unexpanded response continued to the physical cut, as in the
leading-twist construction of Ref.~\cite{Arleo2001}.  In general this
includes both the source-weighted spectral discontinuity and
$\operatorname{Disc}K$ in Eq.~\eqref{eq:boundary-jump}; in the strict
leading-$E1$ Bhanot--Peskin limit the irreducible contribution vanishes
at the retained order.

At short distance, hybrid BO potentials have the form
\begin{equation}
V_{\kappa\lambda}(r)=V_o(r)+\Lambda_\kappa+b_{\kappa\lambda}r^2+\cdots ,
\label{eq:shortBO}
\end{equation}
with a gluelump energy $\Lambda_\kappa$ and channel-dependent corrections~\cite{Berwein2015,BrambillaBO2018}.  
Alongside $r_\phi\Lambda_{\rm QCD}\ll1$, the matching domain must remain separated from source-accessible
hybrid poles and cuts associated with gluelump offsets and
open-flavor thresholds.  Regions of strong level mixing, including
avoided crossings, can likewise invalidate a single-channel local
reduction even though an avoided crossing is not itself a
singularity.  If a source-accessible pole or cut lies in the
low-energy domain, or if strong near-degenerate mixing is present,
the corresponding channels must be retained explicitly through the Feshbach construction of
Section~2.2 rather than absorbed into local Wilson coefficients.  These matching conditions accompany any phenomenological choice of $\epsilon_0$ in the Bhanot--Peskin formula~\cite{Arleo2005,Brambilla2011}.

\section{Coulombic limit}

Throughout this section, $\alpha_s$ denotes $\alpha_s^{\overline{\rm MS}}(\mu_s)$ at a soft scale $\mu_s\sim1/a_0\sim m_Qv$.  The NRQCD coefficient $c_F(\mu_s;\mu_h)$ is matched at $\mu_h\sim m_Q$ and evolved to $\mu_s$, while the pNRQCD one-field coefficients are matched at $\mu_s$.  For the Hermitian EFT of a stable state these coefficients are real; daggers below retain the ordering of their multiplicative $r$ dependence with the resolvent.  At finite $N_c$ the Coulomb potentials are $V_s=-C_F\alpha_s/r$ and $V_o=\alpha_s/(2N_cr)$.  The Bhanot--Peskin limit takes $C_F\to N_c/2$ in the singlet potential and neglects $V_o$, giving
\begin{align}
V_s^{\rm BP}(r)&=-\frac{N_c\alpha_s}{2r},&
\phi_{1S}(r)&=\frac{e^{-r/a_0}}{\sqrt{\pi a_0^3}},\notag\\
E_{1S}&=-\epsilon_0,&
a_0&=\frac{4}{m_QN_c\alpha_s},\qquad
\epsilon_0=\frac{1}{m_Qa_0^2}.
\label{eq:coulomb}
\end{align}
For the strict leading-order Bhanot--Peskin formulas through
Eq.~\eqref{eq:c-to-d}, we set $V_A=1$ and $h_o=\bm p^2/m_Q$, with
$\langle\bm r|\bm p\rangle=e^{i\bm p\cdot\bm r}$ and
$1_o=\int\dd^3p\,|\bm p\rangle\langle\bm p|/(2\pi)^3$. For all free-octet expressions in this section, including the magnetic
formulas below, we therefore use
\[
\Delta_o\equiv h_o-E_{1S}
=\frac{\bm p^2}{m_Q}+\epsilon_0,
\]
unless stated otherwise.  Corrections to the singlet Hamiltonian and
to octet propagation relative to this reference are assigned to the
Hamiltonian-insertion terms introduced below.
The Coulomb wave function, $E1$ overlap, transition density, Bhanot--Peskin cross section, and Mellin moments are established results~\cite{Peskin1979,BhanotPeskin1979,Arleo2001,Brambilla2011}.  In the conventions of Eq.~\eqref{eq:coulomb}, the two inputs needed below are
\begin{equation}
\rho_{1S}^{(r)}(\Delta)\equiv
\sum_i\langle1S|r^i\delta(\Delta-\Delta_o)r^i|1S\rangle
=\frac{2^8}{\pi}\frac{a_0^2}{\epsilon_0}
\frac{(\Delta/\epsilon_0-1)^{3/2}}{(\Delta/\epsilon_0)^6}
\,\theta(\Delta-\epsilon_0).
\label{eq:rhoBP}
\end{equation}
\begin{equation}
\sigma_{\Phi g}^{\rm BP}(\omega)=
\frac{16^3\pi}{6N_c^2}\,a_0^3\epsilon_0
\frac{(\omega/\epsilon_0-1)^{3/2}}{(\omega/\epsilon_0)^5}
\theta(\omega-\epsilon_0).
\label{eq:BP}
\end{equation}
In this projection, the color-weighted source measure introduced in
Sec.~2.3 is
\[
\dd\mu_{ij}(\Delta)
=\frac{T_F}{N_c}
\langle1S|r^i\delta(\Delta-\Delta_o)r^j|1S\rangle
\,\dd\Delta ,
\qquad
\sum_i\frac{\dd\mu_{ii}}{\dd\Delta}
=\frac{T_F}{N_c}\rho_{1S}^{(r)}(\Delta).
\]
With $T_F=1/2$, the standard optical-theorem normalization then relates
the transition density and the cross section by
\begin{equation}
\sigma_{\Phi g}^{\rm BP}(\omega)
=\frac{8\pi^2a_0\epsilon_0\omega}{3N_c^2}\,
\rho_{1S}^{(r)}(\omega).
\label{eq:cross-density}
\end{equation}
The leading-twist moment function is~\cite{Peskin1979,BhanotPeskin1979,Arleo2001}
\begin{equation}
\begin{aligned}
f(x)&=\frac{16^3}{3N_c^2}x^{5/2}(1-x)^{3/2},\\
d_n&=\int_0^1\frac{\dd x}{x}\,x^n f(x)
=\frac{16^3}{3N_c^2}B\!\left(n+\frac52,\frac52\right).
\end{aligned}
\label{eq:dn}
\end{equation}
The single inverse-moment relation required to connect the BOEFT spectral measure to this established normalization is
\begin{equation}
\int_{\epsilon_0}^{\infty}
\frac{\dd\Delta\,\rho_{1S}^{(r)}(\Delta)}{\Delta^{2N-1}}
=
\frac{3N_c^2a_0^2}{16\pi\epsilon_0^{\,2N-1}}\,d_{2N},
\label{eq:inverse-to-dn}
\end{equation}
where $N=1,2,\ldots$.  Isotropy and $a_0\epsilon_0=N_c\alpha_s/4$ then reproduce the standard relation between the two coefficient conventions,
\begin{equation}
\boxed{c_{E,{\rm BP}}^{(N)ij}=\frac{\delta^{ij}}{2}\,d_{2N}}
\label{eq:c-to-d}
\end{equation}
in the strict Bhanot--Peskin, forward on-shell gluon projection~\cite{Peskin1979,BhanotPeskin1979,Arleo2001}.  Within the same Coulombic, free-octet and leading-$E1$ projection, but retaining the multiplicative electric vertex matching, the operator form is
\begin{equation}
\begin{aligned}
c_{E,{\rm E1}}^{(N)ij}={}&
\frac{2\pi\alpha_s\epsilon_0^{\,2N-2}}{N_ca_0^3}
\left\langle1S\left|r^iV_A^\dagger
\Delta_o^{-(2N-1)}V_A r^j\right|1S\right\rangle ,\\[-0.2em]
&\xrightarrow{\,V_A(r;\mu_s)\to V_A(\mu_s)\,}
V_A^2(\mu_s)c_{E,{\rm BP}}^{(N)ij}.
\end{aligned}
\label{eq:matched-electric-coefficient}
\end{equation}
The arrow is the local matching-value approximation used in the ratios below, not an operator identity for an arbitrary $r$-dependent $V_A$.  The complete electric coefficient may be organized as
\begin{equation}
c_E^{(N)ij}=
c_{E,{\rm E1}}^{(N)ij}
+c_{E,\delta H}^{(N)ij}
+c_{E,{\rm 1f,higher}}^{(N)ij}
+c_{E,{\rm 2f,rem}}^{(N)ij}
+\cdots .
\label{eq:complete-electric-coefficient}
\end{equation}
Here $c_{E,{\rm E1}}^{(N)ij}$ is the sequential term in
Eq.~\eqref{eq:matched-electric-coefficient},
$c_{E,\delta H}^{(N)ij}$ collects corrections generated by deviations
of the singlet Hamiltonian and of the octet excitation operator from
the Coulombic free-octet reference, including corrections to the
singlet state and energy and insertions of the repulsive octet
potential,
$c_{E,{\rm 1f,higher}}^{(N)ij}$ contains reducible responses with
higher-order one-field vertices, and
$c_{E,{\rm 2f,rem}}^{(N)ij}$ denotes the remaining irreducible
two-electric-field matching.  In a general matrix element, ${\mathcal Q}_{2N}^E$ decomposes according to Eq.~\eqref{eq:trace-basis}; at the bare matching scale the same short-distance coefficient multiplies its correlated metric-trace components.  The renormalized coefficient vector additionally contains the operators required by mixing, as well as independent magnetic and higher-twist matching terms.

The $M1$ singlet--octet mechanism and its Coulombic gluo-dissociation strength were calculated in Ref.~\cite{ChenHe2017}; the spin-resolved pNRQCD interaction and additional symmetry-allowed one-field tensors at $r^0/m_Q$ were formulated in Ref.~\cite{YangYao2024}.  Write $\mathcal B_i^a\equiv gB_i^a$.  In our tensor basis, define $V_\delta^{(s)}$ as the coefficient of
$\delta_{ij}$ and $V_{\hat r}^{(s)}$ as that of
$\hat r_i\hat r_j$ in the singlet--octet spin vertex.  With the notation
of Ref.~\cite{YangYao2024}, $V_\delta^{(s)}$ corresponds to their
$V_A^s$.  The two equivalent tensor bases are
\begin{equation}
\begin{aligned}
\Gamma_{ij}(\bm r;\mu_s)
&=V_\delta^{(s)}\delta_{ij}+V_{\hat r}^{(s)}\hat r_i\hat r_j\\[-0.2em]
&=V_{\rm iso}^{(s)}\delta_{ij}
+V_T^{(s)}\mathcal T_{ij}(\hat{\bm r}),\qquad
\mathcal T_{ij}=\hat r_i\hat r_j-\frac13\delta_{ij},\\[-0.2em]
V_{\rm iso}^{(s)}&=V_\delta^{(s)}+\frac13V_{\hat r}^{(s)},
\qquad V_T^{(s)}=V_{\hat r}^{(s)} .
\end{aligned}
\label{eq:magnetic-vertex-tensor}
\end{equation}
After the spin projection, the singlet--octet magnetic interaction at this order is
\begin{equation}
\mathcal L_{M1}^{s-o}=\frac{c_F}{m_Q}
\sqrt{\frac{T_F}{N_c}}\left[
S_1^\dagger\mathcal B_i^a\Gamma_{ij}O_{3j}^a
+S_{3j}^\dagger\mathcal B_i^a\Gamma_{ij}O_1^a
+\mathrm{h.c.}\right].
\label{eq:M1-vertex}
\end{equation}
Here $S_1,O_1^a$ and $S_{3i},O_{3i}^a$ are the spin-singlet and Cartesian spin-triplet fields.  The coefficient $c_F$ is denoted $c_4$ in Ref.~\cite{YangYao2024}; we factor it from both tensors by definition.  At tree level, $V_\delta^{(s)}=1$ and $V_{\hat r}^{(s)}=0$; hence $V_{\rm iso}^{(s)}=1+O(\alpha_s(\mu_s))$ and $V_T^{(s)}=O(\alpha_s(\mu_s))$.  Both tensors scale as $r^0/m_Q$ and therefore have the same velocity power.  The additional structures at this order that contain only octet fields do not enter a quadratic elastic singlet response with two one-field insertions.

For a spin-singlet $1S$ state, the sequential response to the rescaled
field $\bm{\mathcal B}=g\bm B$, containing both vertex-matching
structures, is
\begin{align}
\beta_{B,{\rm seq}}^{ij}(\omega)\big|_{{}^1S_0}
={}&\frac{T_Fc_F^2}{N_cm_Q^2}
\left\langle1S\left|
\Gamma_{ik}^\dagger\frac{1}{\Delta_o-\omega-i0}\Gamma_{jk}
+\Gamma_{jk}^\dagger\frac{1}{\Delta_o+\omega+i0}\Gamma_{ik}
\right|1S\right\rangle .
\label{eq:M1-response}
\end{align}
For $V_T^{(s)}=0$ the response is rotationally scalar, and for
$V_{\rm iso}^{(s)}=1$ it reduces to the established leading-order
$M1$ response~\cite{ChenHe2017,YangYao2024}.  When hyperfine effects
in the spatial Hamiltonians are neglected, so that the singlet and
triplet spatial wave functions and resolvents coincide, averaging a
vector ${}^3S_1$ state over its three polarizations multiplies the
scalar-vertex result by $1/3$, while averaging over the complete
$1+3$ hyperfine multiplet multiplies it by $1/2$.  The unaveraged
vector and tensor-vertex responses retain their polarization tensors.

Taking subthreshold inverse moments and using Eq.~\eqref{eq:local-lagrangian} gives the operator form of the reducible sequential coefficient built from the one-field basis through $r^0/m_Q$,
\begin{equation}
c_{B,{\rm seq}}^{(N)ij}\big|_{{}^1S_0}
=\mathcal N_Nc_F^2
\left\langle1S\left|
\Gamma_{ik}^\dagger\Delta_o^{-(2N-1)}\Gamma_{jk}
\right|1S\right\rangle,
\qquad
\mathcal N_N=\frac{2\pi\alpha_s\epsilon_0^{\,2N-2}}
{N_ca_0^3m_Q^2}.
\label{eq:complete-sequential-M1}
\end{equation}
Writing $G_{ij}=V_{\rm iso}^{(s)}\delta_{ij}$ and $T_{ij}=V_T^{(s)}\mathcal T_{ij}$ displays the vertex-matching contributions explicitly:
\begin{align}
c_{B,{\rm seq}}^{(N)ij}={}&c_{B,GG}^{(N)ij}
+c_{B,GT}^{(N)ij}+c_{B,TT}^{(N)ij},\\[-0.2em]
c_{B,GT}^{(N)ij}={}&\mathcal N_Nc_F^2
\langle1S|G_{ik}^\dagger\Delta_o^{-(2N-1)}T_{jk}
+T_{ik}^\dagger\Delta_o^{-(2N-1)}G_{jk}|1S\rangle,
\notag\\[-0.2em]
c_{B,TT}^{(N)ij}={}&\mathcal N_Nc_F^2
\langle1S|T_{ik}^\dagger\Delta_o^{-(2N-1)}T_{jk}|1S\rangle,
\label{eq:vertex-matching-contributions}
\end{align}
with $c_{B,GG}$ obtained by replacing both $T$'s by $G$ in the last line.  All coefficients in this decomposition refer to the spin-singlet
$1S$ response; the state label is suppressed only to simplify the
notation. For a central Coulomb Hamiltonian and an $S$ wave, $c_{B,GT}=0$ because $\mathcal T_{ij}$ carries orbital rank two.  Hence the radiative correction to $V_{\rm iso}^{(s)}$ enters $c_{B,GG}$ at relative $O(\alpha_s)$, whereas the genuinely traceless term starts through $c_{B,TT}$ at relative $O(\alpha_s^2)$.  In noncentral or coupled BO channels the interference need not vanish and is relative $O(\alpha_s)$.

In the same free-octet limit, the isotropic $M1$ response has the
normalized spectral density
\begin{equation}
\rho_{M,1S}^{(0)}(\Delta)\equiv
\langle1S|\delta(\Delta-\Delta_o)|1S\rangle
=
\frac{16}{\pi\epsilon_0}
\frac{\sqrt{\Delta/\epsilon_0-1}}
{(\Delta/\epsilon_0)^4}
\theta(\Delta-\epsilon_0).
\label{eq:M1-spectral-density}
\end{equation}
With the continuum normalization used above,
\begin{equation}
\int_{\epsilon_0}^{\infty}\dd\Delta\,
\rho_{M,1S}^{(0)}(\Delta)=1,
\label{eq:M1-density-normalization}
\end{equation}
and its $(2N-1)$st inverse moment directly gives the first line of
Eq.~\eqref{eq:M1-moments}.
For a local isotropic $V_{\rm iso}^{(s)}$, the positive-frequency
boundary value therefore obeys
\begin{equation}
\operatorname{Im}\beta_{B,GG,+}^{ij}(\omega)
=
\pi\frac{T_Fc_F^2}{N_cm_Q^2}
|V_{\rm iso}^{(s)}|^2
\delta^{ij}\rho_{M,1S}^{(0)}(\omega),
\qquad \omega>0 .
\label{eq:M1-cut}
\end{equation}
Thus the $M1$ cut and the local coefficients below are respectively
the boundary value and inverse moments of the same spectral measure.

For the isotropic component, approximate $V_{\rm iso}^{(s)}(r;\mu_s)$ by its local Coulombic matching value at the characteristic soft scale.  The strict free-octet inverse moment is then
\begin{align}
\left\langle1S\left|\Delta_o^{-(2N-1)}\right|1S\right\rangle
={}&\frac{16}{\pi\epsilon_0^{\,2N-1}}
B\!\left(\frac32,2N+\frac32\right),
\notag\\[-0.2em]
c_{B,GG}^{(N)ij}\big|_{{}^1S_0}
={}&8\alpha_s^2(c_FV_{\rm iso}^{(s)})^2
B\!\left(\frac32,2N+\frac32\right)\delta^{ij},
\label{eq:M1-moments}
\end{align}
where the first line uses $|\widetilde\phi_{1S}(\bm p)|^2=64\pi a_0^3/(1+a_0^2p^2)^4$.  The $N=1$ isotropic coefficient is
\begin{equation}
\boxed{c_{B,GG}^{(1)ij}\big|_{{}^1S_0}
=\frac{5\pi}{16}\alpha_s^2(c_FV_{\rm iso}^{(s)})^2\delta^{ij}},
\qquad
\overline c_{B,GG}^{(1)ij}\big|_{{}^3S_1}=\frac13c_{B,GG}^{(1)ij}\big|_{{}^1S_0}.
\label{eq:first-M1-coefficient}
\end{equation}
The hyperfine-multiplet average is one half of the singlet result at this spin-independent accuracy.  Since $c_{E,{\rm BP}}^{(1)ij}=14\pi\delta^{ij}/(3N_c^2)$, the local matching-value approximation $c_{E,{\rm E1}}^{(1)}=V_A^2(\mu_s)c_{E,{\rm BP}}^{(1)}$ gives the isotropic singlet ratio below, where $V_A\equiv V_A(\mu_s)$.
\begin{equation}
\frac{c_{B,GG}^{(1)}}{c_{E,{\rm E1}}^{(1)}}
=\frac{15}{14}\left(\frac{c_FV_{\rm iso}^{(s)}}{V_A}\right)^2v_C^2,
\qquad v_C=(m_Qa_0)^{-1}=\frac{N_c\alpha_s}{4},
\label{eq:magnetic-electric-ratio}
\end{equation}
which reduces to $(15/14)v_C^2$ only for $c_F=V_{\rm iso}^{(s)}=V_A=1$.  Thus, in this Coulombic reference problem, two $M1$ insertions are
suppressed by $v_C^2$ relative to two $E1$ insertions.  For the $GG$ component displayed in Eq.~\eqref{eq:magnetic-electric-ratio}, radiative matching changes the prefactor through $c_F$, $V_{\rm iso}^{(s)}$, and $V_A$; the full sequential coefficient also contains the $V_T^{(s)}$ terms in Eq.~\eqref{eq:vertex-matching-contributions}.  If an electric or magnetic vertex coefficient is kept as an $r$-dependent operator, the ratio is defined by Eqs.~\eqref{eq:matched-electric-coefficient} and~\eqref{eq:complete-sequential-M1}, not by pulling the coefficients outside the matrix elements.

The one-field response does not exhaust the magnetic coefficient at this order.  The leading covariant kinetic term produces the particle--antiparticle diamagnetic seagull; its kinetic coefficient is fixed to unity by Poincar\'e invariance~\cite{BrambillaPoincare2003}, so no independent multiplicative Wilson coefficient is omitted.  We obtain its QCD color-singlet form by applying the standard conserved-pseudomomentum construction~\cite{AlfordStrickland2013} to an infinitesimal static Cartan background, performing the one-body singlet color projection
$g^2\langle S|T_1^aT_1^b|S\rangle
=(g^2T_F/N_c)\delta^{ab}$, and using global color symmetry to complete the quadratic tensor.  For equal masses the relative orbital Zeeman term vanishes; in the rest frame $\bm K=0$ the motional Stark term also vanishes, while the quadratic term survives:
\begin{equation}
\delta H_{\rm dia}=
\frac{g^2T_F}{4N_cm_Q}
\left(r^2\delta^{ij}-r^ir^j\right)B_i^aB_j^a.
\label{eq:diamagnetic-hamiltonian}
\end{equation}
A derivation from the two covariant kinetic terms is given in
Appendix~\ref{app:diamagnetic}.
Its contribution to Eq.~\eqref{eq:local-lagrangian} has the opposite sign because $\delta E=-\delta\mathcal L_{\rm pol}$:
\begin{equation}
c_{B,{\rm dia}}^{(1)ij}=
-\frac{g^2T_F}{4N_cm_Qa_0^3}
\left\langle r^2\delta^{ij}-r^ir^j\right\rangle.
\label{eq:diamagnetic-coefficient}
\end{equation}
For the Coulombic $1S$ state, $\langle r^ir^j\rangle=a_0^2\delta^{ij}$ and $\langle r^2\rangle=3a_0^2$, so
\begin{equation}
\boxed{c_{B,{\rm dia}}^{(1)ij}\big|_{1S}
=-\frac{\pi}{4}\alpha_s^2\delta^{ij}},
\qquad
\frac{c_{B,{\rm dia}}^{(1)}}{c_{E,{\rm E1}}^{(1)}}=-\frac{6}{7V_A^2}v_C^2 .
\label{eq:first-diamagnetic-coefficient}
\end{equation}
This spin-independent term is the same for the singlet and triplet.  For a spin-singlet $1S$ state, combining the two calculated pieces while retaining the multiplicative one-field matching factors gives the partial ratio
\begin{equation}
\frac{c_{B,GG}^{(1)}+c_{B,{\rm dia}}^{(1)}}{c_{E,{\rm E1}}^{(1)}}
=\frac{15(c_FV_{\rm iso}^{(s)})^2-12}{14V_A^2}\,v_C^2.
\label{eq:matched-partial-magnetic-ratio}
\end{equation}
At strict tree level, $c_F=V_{\rm iso}^{(s)}=V_A=1$, this becomes
\begin{equation}
\frac{c_{B,GG}^{(1)}+c_{B,{\rm dia}}^{(1)}}{c_{E,{\rm BP}}^{(1)}}
=\frac{3}{14}v_C^2,
\label{eq:tree-leading-magnetic-ratio}
\end{equation}
before the remaining Hamiltonian, higher-one-field, and irreducible two-field terms are added.

Equation~\eqref{eq:complete-sequential-M1} is the complete reducible
two-insertion response generated by the stated pNRQCD one-field basis
through $r^0/m_Q$, evaluated with the leading reference Hamiltonian.
In the following decomposition
$c_{B,{\rm seq}}^{(1)}[\Gamma]$ denotes this reference-Hamiltonian
contribution, while $c_{B,\delta H}^{(1)}$ collects corrections
generated by Hamiltonian insertions.  The dimension-four coefficient also contains the leading diamagnetic part of the irreducible two-field kernel, sequential terms from higher-order one-field vertices, and the remaining local matching corrections,
\begin{equation}
c_B^{(1)}=c_{B,{\rm seq}}^{(1)}[\Gamma]
+c_{B,{\rm dia}}^{(1)}+c_{B,\delta H}^{(1)}
+c_{B,{\rm 1f,higher}}^{(1)}+c_{B,{\rm 2f,rem}}^{(1)}+\cdots .
\label{eq:complete-magnetic-coefficient}
\end{equation}
Here $c_{B,{\rm 1f,higher}}^{(1)}$ collects reducible responses containing at least one higher-$1/m_Q$ one-field vertex.  The term $c_{B,{\rm 2f,rem}}^{(1)}$ is the irreducible two-field matching remainder, including hard local $B^2/m_Q^3$ matching, after the explicitly separated diamagnetic term.  These definitions prevent double counting in Eq.~\eqref{eq:complete-magnetic-coefficient}.  In Coulombic counting, both $c_{B,{\rm seq}}^{(1)}/c_E^{(1)}$ and $c_{B,{\rm dia}}^{(1)}/c_E^{(1)}$ are $O(v^2)$.  Spin-dependent or relativistic Hamiltonian insertions satisfy $\delta H/\Delta_o=O(v^2)$; higher-$1/m_Q$ one-field terms and hard local $B^2/m_Q^3$ matching are likewise suppressed relative to these leading magnetic pieces, absent an additional infrared enhancement.  Matching loops instead supply powers of $\alpha_s$ at fixed velocity order.  Beyond the weak-coupling reference problem, nonperturbative BO dynamics
enters both through the channel-complete complementary-sector resolvent
and the matched one-field source maps, and through the irreducible
two-field kernel.  These contributions need not obey the Coulombic estimate.  Higher multipoles with additional background gradients populate separate higher-dimension operators.  For the isotropic projection, the ratio $\alpha_B^\Phi/\alpha_E^\Phi=c_B^{(1)}/c_E^{(1)}$ in the convention of Eq.~\eqref{eq:physical-polarizabilities} involves the full matched coefficients, not merely the partial ratio in Eq.~\eqref{eq:matched-partial-magnetic-ratio}~\cite{ScalarSpinTwo2026}.
The nonoverlapping bookkeeping of the magnetic coefficient is shown
schematically in Fig.~\ref{fig:magnetic-bookkeeping}.
\begin{figure}[t]
\centering
\begin{tikzpicture}[
  font=\small,
  >=Latex,
  state/.style={draw,rounded corners=2pt,minimum width=8mm,
    minimum height=6mm,fill=gray!8},
  vertex/.style={circle,draw,thick,minimum size=6mm,inner sep=0pt,
    fill=white},
  resolvent/.style={draw,thick,rounded corners=2pt,minimum width=25mm,
    minimum height=8mm,fill=blue!8},
  contact/.style={diamond,draw,thick,aspect=1.8,minimum width=13mm,
    minimum height=8mm,fill=orange!12},
  remainder/.style={draw,thick,rounded corners=2pt,minimum width=30mm,
    minimum height=9mm,fill=green!9,align=center},
  field/.style={->,thick,teal!65!black},
  prop/.style={->,thick}
]
\node[anchor=west,font=\bfseries] at (0,2.30)
  {(a) Reducible sequential term};

\node[state] (sL) at (0.65,1.25) {$\Phi$};
\node[vertex] (mL) at (2.00,1.25) {$M1$};
\node[resolvent] (oR) at (4.20,1.25) {$\Delta_o^{-1}$};
\node[vertex] (mR) at (6.40,1.25) {$M1$};
\node[state] (sR) at (7.75,1.25) {$\Phi$};

\draw[prop] (sL)--(mL);
\draw[prop] (mL)--(oR);
\draw[prop] (oR)--(mR);
\draw[prop] (mR)--(sR);

\draw[field] (2.00,2.05)--node[right] {$\mathcal B_j^b$} (mL);
\draw[field] (6.40,2.05)--node[right] {$\mathcal B_i^a$} (mR);

\node[anchor=west,align=left] at (8.55,1.25)
  {$c_{B,{\rm seq}}^{(1)}[\Gamma]$\\
   (two one-field vertices)};

\node[anchor=west,font=\bfseries] at (0,0.15)
  {(b) Irreducible kinetic seagull};

\node[state] (dL) at (0.65,-1.15) {$\Phi$};
\node[contact] (dia) at (4.20,-1.15) {$B^2$};
\node[state] (dR) at (7.75,-1.15) {$\Phi$};

\draw[prop] (dL)--(dia);
\draw[prop] (dia)--(dR);

\draw[field] (3.70,-0.20)--node[left] {$B_j^b$} (dia.north west);
\draw[field] (4.70,-0.20)--node[right] {$B_i^a$} (dia.north east);

\node[anchor=west,align=left] at (8.55,-1.15)
  {$c_{B,{\rm dia}}^{(1)}$\\
   (local two-field vertex)};

\draw[gray!45] (0,-2.00)--(15.5,-2.00);

\node[anchor=west,font=\bfseries] at (0,-2.50)
  {(c) Additional matched contributions};

\node[remainder] (dh) at (1.80,-3.55)
  {Hamiltonian insertions\\$c_{B,\delta H}^{(1)}$};
\node[remainder] (hf) at (6.15,-3.55)
  {higher one-field vertices\\$c_{B,{\rm 1f,higher}}^{(1)}$};
\node[remainder] (tf) at (10.70,-3.55)
  {two-field remainder\\$c_{B,{\rm 2f,rem}}^{(1)}$};

\node[anchor=west] at (12.60,-3.55) {$+\;\cdots$};

\node[
  draw,
  rounded corners=3pt,
  fit=(dh)(hf)(tf),
  inner sep=5pt,
  dashed,
  gray!70
] {};
\end{tikzpicture}
\caption{Nonoverlapping organization of the dimension-four magnetic
matching coefficient.  The reducible two-$M1$ response, the local
covariant-kinetic seagull, Hamiltonian insertions, higher-order
one-field sources, and the remaining irreducible two-field matching
are assigned to distinct terms in
Eq.~\eqref{eq:complete-magnetic-coefficient}.  Only the sequential
leading-$M1$ term and the diamagnetic seagull are evaluated explicitly
in the Coulombic approximation used here.  The internal line denotes
the appropriate octet resolvent rather than a free perturbative
propagator.}
\label{fig:magnetic-bookkeeping}
\end{figure}

In this normalization Eq.~\eqref{eq:BP} is proportional to the
positive-frequency boundary-value jump, while Eq.~\eqref{eq:dn}
contains the inverse moments of the same projected resolvent.  In the
strict free-octet BP projection, the local resolvent series converges
for $|\omega|<\epsilon_0$, the distance to the open-octet threshold.
For the full matched response, the convergence radius is instead set
by the nearest source-accessible pole or cut and by the analyticity
domain of the irreducible kernel $K$, as discussed in Sec.~2.3.
Above threshold the unexpanded response is defined by its integral
representation and boundary value on the cut; in the general matched
response this may also include $\operatorname{Disc}K$.

A finite-$N_c$ Coulombic treatment uses
$V_s=-C_F\alpha_s/r$ together with the repulsive octet potential
$V_o=\alpha_s/(2N_cr)$.  If these potentials are incorporated into
the reference problem, the free-octet plane wave is replaced by the
octet scattering state, producing in particular the finite-$N_c$
final-state-interaction correction of Ref.~\cite{Brambilla2011}.
In the free-octet reference convention adopted above, the same
deviations are instead assigned to
$c_{E,\delta H}^{(N)}$ and $c_{B,\delta H}^{(N)}$ and must not also be
included in the sequential reference terms.  The strict
Bhanot--Peskin formula is therefore a further limit of the
weak-coupling resolvent response.

\section{Conclusions}

We have formulated the source-dependent gluonic response on the
zero-field BOEFT Hilbert space.  Matching normalized source residues
and their directional derivatives defines the transition vectors and,
after subtraction of the resolvent-reducible contribution, the
irreducible two-field kernel.  The temporal adjoint line, covariant
inner product, and singlet map connect the auxiliary color fiber to
physical BOEFT states.  The Schur complement separates channels retained dynamically from
the eliminated sector.  When the eliminated $Q_H$ block is
spectrally separated from the matching domain, its $G_H$ dependence
admits a local expansion, while the final retained-sector resolvent
continues to carry the poles and cuts of the channels kept
dynamically.

Multipole and source-resolvent locality, analyticity of the
irreducible kernel $K_{ij}$ in the matching domain, and the joint
propagation--vertex bounds in
Eqs.~\eqref{eq:factorization-bound}--\eqref{eq:factorization-error}
provide sufficient criteria for a channel-factorized local OPE.
These bounds do not by themselves establish that $q_H$ and $q_d$ are
small; their smallness must follow from the relevant EFT power
counting.  Weak-coupling pNRQCD calculability additionally requires
Eq.~\eqref{eq:weak-coupling-conditions}, while the combined locality,
factorization, source-gap, and weak-coupling conditions are summarized
in Eq.~\eqref{eq:hierarchy-summary}.  Identification with the pure
Peskin inverse moments further requires the irreducible kernel to
vanish at the retained order and uses the leading-$E1$ choice
$V_A=1$ together with the large-$N_c$/free-octet reference problem.

At the leading matching order displayed here, the strict
four-dimensional harmonic decomposition of each untraced electric
operator produces correlated twist-two and metric-trace components.
A dimensionally regulated loop treatment requires the corresponding
enlarged $d$-dimensional operator basis, and renormalization need not
preserve a common coefficient for the different sectors.  In the Coulombic $1S$,
leading-$E1$, free-octet, and forward on-shell gluon projection, the
established Bhanot--Peskin relation
\[
c_{E,{\rm BP}}^{(N)ij}=\frac{\delta^{ij}}{2}d_{2N}
\]
follows, while the positive-frequency boundary-value jump of the same
spectral measure reproduces the Bhanot--Peskin dissociation kernel.
Retaining electric vertex matching gives the ordered operator
expression in Eq.~\eqref{eq:matched-electric-coefficient}, which
reduces to
$V_A^2(\mu_s)c_{E,{\rm BP}}^{(N)ij}$ only in the local
matching-value approximation.

For a spin-singlet Coulombic $1S$ state, in the free-octet and local
matching-value approximations, the isotropic $GG$ component of the
sequential $M1$ response gives
\[
c_{B,GG}^{(1)ij}
=\frac{5\pi}{16}\alpha_s^2
(c_FV_{\rm iso}^{(s)})^2\delta^{ij}.
\]
When hyperfine effects in the spatial Hamiltonians are neglected, the
polarization-averaged vector and hyperfine-multiplet averages of this
scalar-vertex contribution are respectively $1/3$ and $1/2$ of the
spin-singlet result.  The tensor decomposition in
Eq.~\eqref{eq:magnetic-vertex-tensor} organizes the additional
radiative $GT$ and $TT$ contributions; the $GT$ term vanishes for a
central Coulomb Hamiltonian and an $S$ wave but need not vanish in
noncentral or coupled BO channels.

At the same relative $O(v^2)$ order in Coulombic counting, the
spin-independent covariant-kinetic seagull gives
\[
c_{B,{\rm dia}}^{(1)ij}
=-\frac{\pi}{4}\alpha_s^2\delta^{ij}.
\]
The two explicitly calculated magnetic pieces give the partial ratio
in Eq.~\eqref{eq:matched-partial-magnetic-ratio}.  At strict tree
level,
$c_F=V_{\rm iso}^{(s)}=V_A=1$, this partial benchmark becomes
\[
\frac{c_{B,GG}^{(1)}+c_{B,{\rm dia}}^{(1)}}
     {c_{E,{\rm BP}}^{(1)}}
=\frac{3}{14}v_C^2
\]
for the spin singlet.  It is not the complete physical magnetic-to-
electric ratio: in the isotropic projection,
$\alpha_B^\Phi/\alpha_E^\Phi=c_B^{(1)}/c_E^{(1)}$ must be formed from
the full matched coefficients in
Eqs.~\eqref{eq:complete-electric-coefficient} and
\eqref{eq:complete-magnetic-coefficient}.  The construction therefore
separates the resolvent contributions fixed by the transition sources
from Hamiltonian, higher-vertex, and irreducible two-field corrections
that require independent matching.

\appendix
\section{Covariant-kinetic seagull in a constant Cartan background}
\label{app:diamagnetic}

Consider an infinitesimal homogeneous background along one Cartan
generator and denote the corresponding Abelianized heavy-quark charge
by $q$.  In the symmetric gauge
$\bm A(\bm x)=\bm B\times\bm x/2$, the equal-mass kinetic Hamiltonian is
\begin{equation}
H_{\rm kin}=
\frac{[\bm p_1-q\bm A(\bm r_1)]^2}{2m_Q}
+\frac{[\bm p_2+q\bm A(\bm r_2)]^2}{2m_Q}.
\label{eq:two-body-kinetic-B}
\end{equation}
A translationally invariant interaction $V(r)$ may be added throughout;
it is unaffected by the pseudoseparation and is suppressed here because
only the field-dependent kinetic terms are required.
With
\[
\bm R=\frac{\bm r_1+\bm r_2}{2},\qquad
\bm r=\bm r_1-\bm r_2,\qquad
\bm P=\bm p_1+\bm p_2,\qquad
\bm p=\frac{\bm p_1-\bm p_2}{2},
\]
the conserved pseudomomentum is
\begin{equation}
\bm K=\bm p_1+\bm p_2+\frac{q}{2}\bm B\times\bm r .
\label{eq:pseudomomentum}
\end{equation}
Before fixing the pseudomomentum, Eq.~\eqref{eq:two-body-kinetic-B}
can be rewritten as
\[
H_{\rm kin}
=
\frac{\left(\bm P-\frac q2\bm B\times\bm r\right)^2}{4m_Q}
+
\frac{\left(\bm p-\frac q2\bm B\times\bm R\right)^2}{m_Q}.
\]
For an eigenstate of $\bm K$, write
\[
\Psi_{\bm K}(\bm R,\bm r)
=
\exp\!\left[
i\left(\bm K-\frac q2\bm B\times\bm r\right)\cdot\bm R
\right]\psi_{\bm K}(\bm r).
\]
The two covariant momentum combinations then act as
\[
\begin{aligned}
\left(\bm P-\frac q2\bm B\times\bm r\right)\Psi_{\bm K}
&=
e^{i(\cdots)}
\left(\bm K-q\bm B\times\bm r\right)\psi_{\bm K},\\
\left(\bm p-\frac q2\bm B\times\bm R\right)\Psi_{\bm K}
&=
e^{i(\cdots)}\bm p\,\psi_{\bm K}.
\end{aligned}
\]
Consequently, the standard pseudoseparation
of Ref.~\cite{AlfordStrickland2013} gives
\begin{equation}
H_{\rm kin}=
\frac{\bm K^2}{4m_Q}
+\frac{\bm p^2}{m_Q}
-\frac{q}{2m_Q}(\bm K\times\bm B)\!\cdot\!\bm r
+\frac{q^2}{4m_Q}(\bm B\times\bm r)^2 .
\label{eq:pseudoseparated-kinetic}
\end{equation}
The relative orbital Zeeman term vanishes for equal masses.  In the zero-pseudomomentum sector, $\bm K=0$, which is the appropriate
rest-sector choice for the neutral pair, the motional Stark term also
vanishes, leaving
\begin{equation}
\delta H_{\rm dia}=
\frac{q^2}{4m_Q}
(r^2\delta^{ij}-r^ir^j)B_iB_j .
\label{eq:abelian-diamagnetic}
\end{equation}
For the generator acting on either constituent of a normalized
color-singlet heavy pair,
\begin{equation}
\langle S|T_1^aT_1^b|S\rangle
=
\langle S|T_2^aT_2^b|S\rangle
=
\frac{T_F}{N_c}\delta^{ab},
\label{eq:singlet-one-body-color}
\end{equation}
where $T_2^a$ is understood in the antiquark representation.
Thus the quadratic one-body charge factor in the Cartan calculation
is promoted according to
\[
q^2B_iB_j\;\longrightarrow\;
g^2\langle S|T_1^aT_1^b|S\rangle B_i^aB_j^b
=
\frac{g^2T_F}{N_c}B_i^aB_j^a .
\] 
No additional factor of two is required: the coefficient
$q^2/(4m_Q)$ in Eq.~\eqref{eq:abelian-diamagnetic} already results
from the sum of the quark and antiquark kinetic terms.  Equivalently,
on the singlet $T_2^a|S\rangle=-T_1^a|S\rangle$, so the opposite
constituent charges have already been incorporated in the neutral-pair
pseudoseparation. This promotion determines the color-singlet quadratic response about
zero background.  It does not assume the existence of a conserved
pseudomomentum for a general finite non-Abelian background.
Global color invariance then gives
Eq.~\eqref{eq:diamagnetic-hamiltonian}.  The opposite sign of the
corresponding local Lagrangian term follows from
$\delta E=-\delta\mathcal L_{\rm pol}$. Combining this sign with the normalization in
Eq.~\eqref{eq:local-lagrangian} gives
Eq.~\eqref{eq:diamagnetic-coefficient}.  For the BP Coulombic $1S$
state,
$\langle r^ir^j\rangle=a_0^2\delta^{ij}$ and
$\langle r^2\rangle=3a_0^2$; using
$g^2=4\pi\alpha_s$, $T_F=1/2$, and
$m_Qa_0=4/(N_c\alpha_s)$ then reproduces
\[
c_{B,{\rm dia}}^{(1)ij}
=-\frac{\pi}{4}\alpha_s^2\delta^{ij}.
\]

\end{document}